\documentclass[english,12pt]{article}
\usepackage[T2A]{fontenc}
\usepackage{babel}
\usepackage {epsfig}
\usepackage{graphicx,epsfig}
\usepackage{a4}
\usepackage{epsfig}
\def\fun#1#2{\lower3.6pt\vbox{\baselineskip0pt\lineskip.9pt
\ialign{$\mathsurround=0pt#1\hfil##\hfil$\crcr#2\crcr\sim\crcr}}}
\usepackage{amsmath}
\usepackage{amssymb}
\usepackage{graphicx}

\def\fun#1#2{\lower3.6pt\vbox{\baselineskip0pt\lineskip.9pt
\ialign{$\mathsurround=0pt#1\hfil##\hfil$\crcr#2\crcr\sim\crcr}}}

\begin{document}

\hspace*{14cm}{\Large {\it DRAFT}}\\

\begin{center}
\vspace{-0.7cm} {\Large \textbf{CPT invariance and neutrino
physics}} \setcounter{footnote}{-1} \footnote{The early version of
this article has been published in the Russian journal \emph{Uspekhi
Fizicheskikh Nauk}:  $UFN$, \textbf{175}, 863 (2005).  Although the
English translation in \emph{Physics--Uspekhi}, \textbf{48}, 825
(2005) was performed by non-experts, the author has not got
possibility to check up proofs carefully because of strict time
schedule imposed by  the editorial board. Now we present a revised and updated article in English.}\\[6pt]

{I. S. Tsukerman} \\

\it {Institute of Theoretical and Experimental Physics, 117218,
Moscow, Russian Federation}\\E-mail: zuckerma@itep.ru\\

\end{center}
 \textbf{Abstract}

 CPT invariance in neutrino physics has attracted
attention after the revival of the hypothetical idea that neutrino
and antineutrino might  have nonequal masses ($m_{\bar\nu}  \neq
m_{\nu}$) when realizing neutrino oscillations as a new sensitive
phenomenon to search for the violation of this
 fundamental symmetry. Moreover, the profound relation between the CPT and Lorentz symmetries turns the studies
 of CPT and Lorentz invariance violations into the {\bf one two-sided} problem.
 We present a  guide for non-experts through the literature on neutrino physics.  The basic works are reviewed
 thoroughly while for the other papers only current results or discussion issues are quoted. The review covers,
 mostly, oscillations of neutrinos, resonant change of their flavors and cosmic neutrino physics to systematize
 possible evidences of CPT/Lorentz violation in this  sector of the Standard
 Model.\\

\begin{center}
{\large\bf{\it Contents }}\\
\end{center}

\noindent\vspace*{0.15cm}\textbf{Introduction}\\
\textbf{1.  The CPT theorem (quotations)}\\
\textbf{2.  Theoretical and experimental status of the CPT }\\
\vspace*{-0.15cm}\textbf{3.  General consequences of hypothetical CPT and Lorentz invariance violation \\
\hspace*{0.7cm}in neutrino physics}\\
\hspace*{0.5cm} A. CPT and neutrino physics\\
\hspace*{0.5cm}\vspace*{-0.15cm} B.\; Extension of the Standard Model: spontaneous violation \\
\hspace*{1.3cm}of Lorentz invariance and СPТ \\
\hspace*{0.5cm}\vspace*{-0.15cm} C.\; Perturbation-theory formalism for violation of Lorentz invariance, CPT \\
\hspace*{1.3cm}and equivalence principle\\
\hspace*{0.5cm}\vspace*{-0.15cm} D.\;`Non-standard'\, violating mechanisms (quantum decoherence,\\
\hspace*{1.3cm}modified dispersion relations)\\
\vspace*{-0.15cm}\textbf{4. Experimental and observational consequences of CPT and Lorentz invariance\\
\hspace*{0.7cm}violation in neutrino physics}\\
\hspace*{0.5cm} E. Perturbative violation of Lorentz invariance and CPT\\
\hspace*{0.5cm}\vspace*{-0.15cm} F. Violation of Lorentz invariance and equivalence principle\\
\hspace*{1.3cm}in terrestrial and cosmic neutrino physics \\
\hspace*{0.5cm}\vspace*{-0.15cm} G. Manifestations of Lorentz invariance and CPT with `non-standard'\, mechanisms\\
\hspace*{1.3cm}in neutrino physics\\
\textbf{5.  Interpretations of neutrino oscillations based on CPT violation}\\
\hspace*{0.5cm} H. False  CPT-odd effects in matter\\
\hspace*{0.6cm} I. Neutrino oscillations due to violation of Lorentz invariance\\
\hspace*{0.5cm} J. CPT non-invariant `ether', quantum decoherence, and  LSND anomaly \\
\hspace*{0.5cm}\vspace*{-0.15cm} K. Antineutrino \textsl{vs} neutrino oscillation parameters;\\
\hspace*{1.3cm}LSND-anomaly models with \textsl{m$_{\bar \nu}\neq ~$m$_{\nu}$}\\
\textbf{Conclusion}\\
\textbf{References}\\
\newpage
\noindent\textbf{\large Introduction}\\[-10pt]

This review \setcounter{footnote}{0} involves more than two hundred publications
that study implications of the CPT violation
in various neutrino processes and analyze experimental data on the oscillations of solar, atmospheric and reactor
 neutrinos and antineutrinos, including the well-known LSND anomaly. However for ten years past,  among
 review papers and talks on neutrino oscillations (see Refs. \cite{leu} -- \cite{DO}) solely Leung's talk \cite{leu},
 published in 2000, was devoted in total to summarizing neutrino tests of general and special relativity
 while only a few ones touched on the CPT symmetry or treated the question in more details.
  Talks of Akhmedov \cite{akh} and of Kayser \cite{kai} at the `Neutrino-2002'\, and the plenary talk of
Gonzalez-Garcia \cite{G} at the `ICHEP-2002'\, contained brief remarks on $m_{\bar \nu} \neq m_\nu$.
Before the `Neutrino 2004', three Mavromatos' review works on this theme  (talks \cite{MAVR,MA} and
lectures \cite{ma}) were published. Almost at that time  the analytical review by Bahcall, Gonzalez-Garcia and
Pe\~{n}a-Garay \cite{BGP} with  a relevant section  appeared.
 And  at the `Neutrino-2004' de Gouv\^{e}a  included two sections on tests of Lorentz and CPT invariance in
neutrino physics into his  talk \cite{deG}. Since then, the authors
of many survey works are discussing these problems. \footnote{In
this connection, see also section "CP and CPT violation"\, in  very detailed and informative Web-site
created by C.\,Giunti \cite{Giunti}.}

Several remarks are due on the structure of the present paper and
the notation used. By updating the 2004 version of this review the number of referred works was
approximately doubled, so that the reference list covers now about
ten years (up to January, 2009 deadline). The main body of the review is
presented in five Sections while Introduction has a rather
informative   character; in Conclusion one makes quotations from
the few published Solomonian-type outlooks we know.\\
 Publications devoted to general theoretical status of CPT and Lorentz invariance are
given in Sections 1 -- 2. Concrete substantial issues of hypothetical CPT violation are distributed  among eleven
alphabetical Subsections A -- K.   Section 3 (Subsections A, B, C) devoted to general basic works:
 while Subsection A is of introductory character, Subsection B, C, and D describe, what type of CPT/LI-violating
  parameters may be introduced, in principle, into the lagrangian formalism for NO. Sections 4 (Subsections E, F, G)
  and 5 (Subsections H, I, J, K) give to the reader
  up-to-date information about experimental and observational bounds on these parameters in the framework of the
  theoretical approaches.\\  Note here that while the relevant phenomenological methods made used in Subsections B and C
 are a natural generalization of the Standard Model, the `unconventional'\,
 ones discussed in Subsection D have typically problematic validity.\\
To compensate some ambiguity when choosing Subsections for reviewing a given work, we  use cross references and footnotes.
Numerous remarks beyond topics discussed, which we consider as a supplement information, are given as footnotes too.\\
We used the following   abbreviations: CPT, NO, MSW, LI, EP, QD  and MDR;
these stand for, respectively,  the CPT symmetry, neutrino oscillations, the
Mikheyev--Smirnov--Wolfenstein resonance solution in  medium, the
Lorentz invariance, the equivalence principle, the quantum decoherence, and the modified dispersion
relations (considered as relations between energy, $E$, 3-momentum, $p$, and mass, $m$, of a particle).\\[-5pt]

\noindent\textbf{\large 1. The CPT theorem (quotations)}\\[-10pt]

 The theoretical basing
of the Pauli--L\"{u}ders--Schwinger CPT theorem is not the goal
of this review. So, in this Section we present the fundamental
conclusions on CPT symmetry mostly by quotations from familiar
monographs and collected volumes (see also Section 2).\\[-15pt]

$\bullet$ ... Unter sehr allgemeinen und wohlbegr\"{u}ndeten
Voraussetzungen, zu denen die fur die spezielle
Relativit\"{a}tstheorie characteristische Lorentz-Invarianz
geh\"{o}rt, gilt n\"{a}mlich das sogenannte CPT-Theorem. Dieses sagt
aus, dass aus diesen allgemeinen Voraussetzungen -- wir verweisen
f\"{u}r Einzelheiten hier auf die Literatur \footnote{Das
CPT-Theorem wurde zuerst von G.\,L\"{u}ders \cite{lu1} klar erkannt.
-- Ferner: J.\,Schwinger \cite{sch}. -- W.\,Pauli \cite{p1}. --
F\"{u}r nicht lokale Theorien gab R.\,Jost \cite{j1} eine dem
CPT-Theorem \"{a}quivalente Bedingung, die f\"{u}r lokale Theorien
identisch erf\"{u}llt ist. -- Weitere Anwendungen s. T.\,D.\,Lee,
R.\,Oehme und C.\,N.\,Yang \cite{lee}. (Die Fu{\ss}note von Pauli \cite{p2}, \S\,I)}
-- die Invarianz der Theorie f\"{u}r die Zusammensetzung
(Produkt) aller drei Operationen C, P und T (in irgend einer Reihenfolge) bereits folgt.\\
 Dieses hat unter anderem zur Folge, dass die Massen von Teilchen
 und Antiteilchen (allgemeiner die Energiewerte eines Systems von
 Teilchen und die der zu ihnen C-konjugierten Teilchen) einander
  gleich sein m\"{u}ssen. (W.\,Pauli \cite{p2}, \S 1) \\
 \{... Very general and well-founded assumptions, including the
 requirement of Lorentz invariance in special relativity, imply the
 so-called CPT theorem. The theorem states that these general
 assumptions (for details see the literature \footnote{In the first time,
CPT-Theorem was realized by G.\,L\"{u}ders \cite{lu1}.
See, besides: J.\,Schwinger \cite{sch}; W.\,Pauli \cite{p1}.
 For nonlocal theories R.\,Jost \cite{j1} produced a condition equivalent to the CPT theorem that
holds identically for local theories. For further applications see:
T.\,D.\,Lee, R.\,Oehme and C.\,N.\,Yang \cite{lee}. (Pauli's footnote \cite{p2}, \S\,I)})  immediately
 imply the invariance of the theory relative to the combined action (product)
 of all three operations C, P, T (in an arbitrary order). This in
 turn implies, among other things, that the masses of particles and
 antiparticles (in the general case -- energy levels for two systems: of
the particles and of their charge-conjugated particles) must be
strictly identical. (W.\,Pauli \cite{p2}, \S 1)\}\\[-15pt]

 $\bullet$ ... We assume
further for the sake of simplicity the {\it local} character of the
field equation, which means that all field quantities are spinors or
tensors of finite rank and that the interaction part of the
Lagrangian (or the Hamiltonian) contains only derivatives of
finite order of these field quantities... (W.\,Pauli \cite{p1}, \S 1)\\[-15pt]

 $\bullet$  Unabh\"{a}ngig von Schwinger  \cite{sch} kam L\"{u}ders \cite{lu1}
zu dem sehr nahverwandten Resultat, dass unter sehr weiten
Voraussetzungen eine P invariante Theorie, in welcher die normalen
Vertauschungsrelationen bestehen, automatisch CT invariant ist.\\
Die endg\"{u}ltige und allgemeine Formulierung des hier zust\"{a}ndigen
Theorems aber stammt wiederum von Pauli \cite{p1} und lautet {\it CTP
 Theorem}: Eine bez\"{u}glich der eigentlichen Lorentzgruppe invariante
Feldtheorie mit normalen Vertauschungsrelationen ist auch CTP invariant. \\
Der Fortschritt der neuen Fassung besteht darin, dass
(nat\"{u}rlich vor der Entdeckung der Parit\"{a}tsverletzung) nur die
Invarianz bez\"{u}glich der eigentlichen Lorentzgruppe vorausgesetzt
wird. Ausserdem wird das Theorem f\"{u}r beliebigen Spin bewiesen,
w\"{a}hrend L\"{u}ders sich auf die wichtigsten Spinwerte 0, 1/2 und 1
beschr\"{a}nkt.(R.\,Jost \cite{j2}, Abschn. 1, \S 3)\\
\{L\"{u}ders, independently of Schwinger \cite{sch}, obtained a very similar result \cite{lu1}, namely
 that under not very restricting assumptions, a P-invariant theory
 with {\it normal} commutation relations is automatically invariant
 under СТ. \\
 However, the final formulation of this theorem is again
 Pauli's \cite{p1}; the CPT theorem states: a field theory with normal
 commutation relations, invariant under the Lorentz eigengroup, is
 also СРТ-invariant.\\
  The advantage of the new formulation is the
 fact that only invariance under Lorentz eigengroup is assumed
 (obviously, prior to the discovery of parity non-conservation).
 Furthermore, the theorem is proved for an arbitrary spin while
 L\"{u}ders only considered the more important spin values 0, 1/2 and 1.
 (R.\,Jost \cite{j2}, Ch. 1, \S 3)\}\\[-15pt]

$\bullet$ Normal commutation
relations are defined as follows: tensor fields (belonging to
one-valued representations of $L^\uparrow_+$) commute with
themselves and with the spinor fields (belonging to two-valued
representations of $L^\uparrow_+$) at space-like separation;
spinor fields anticommute at space-like separation...\\
 If we anticipate the results of the last chapter, where particles are
introduced into a Wightman field theory, then the above results
imply the law of the connection between spin and statistics:
particles with integer spin obey Bose--Einstein statistics,
particles with half integer spin obey Fermi--Dirac statistics.
(R.\,Jost \cite{j3}, Ch. V, \S 3)\\[-15pt]

$\bullet$ Let us next consider the restrictions imposed by the
requirement that the theory be invariant under (Wigner) time
inversion. An important theorem due to Pauli \cite{p1} and
L\"{u}ders \cite{lu1} (this discovery was essentially anticipated by
Shell (1948) and by Schwinger \cite{Schwi}), and currently known as
the TCP theorem, asserts that within the framework of
relativistically invariant \emph{local} field theories, assuming the
usual connection between spin and statistics, invariance under time
reversal is equivalent to invariance under $U_P U_C$, i.e., the
combined operation of charge conjugation ($U_C$) and space inversion
($U_P$). In a Lagrangian formulation, the TCP theorem is a result of
the assumed invariance under proper Lorentz transformation of
$\mathcal{L}$ [Lagrangian density], the hermiticity of $\mathcal{L}$ , the
locality of the theory, and the assumption that particles of
integer spin (bosons) must obey Bose--Einstein statistics and those
of half-integer spin (fermions) must obey Fermi--Dirac statistics,
i.e., the particles obey the usual
connection with statistics. (\cite{Schwe}, Ch. 10, \S 2, p. 264)\\[-15pt]

$\bullet$ ...Hence it follows that the Lagrangian (14.16) from which
we have demanded only that it must be Hermitian and invariant under
the proper Lorentz transformations, is also  invariant under РСТ
(CPT, TCP, and so on). This is the essence of the L\"{u}ders--Pauli
CPT theorem (for further  details  see Pauli \cite{p1} and
Gravert, L\"{u}ders and Rollnik \cite{GLR})... \\
The requirement that the interaction is local has played an
essential role in the above discussion. In the axiomatic
formulation of quantum field theory, this requirement can be made
less stringent. The proof of the CPT theorem in  axiomatic
approach has been given by  Jost  \cite{j3}, Streater and Wightman
\cite{cpt}, and  Bogolyubov, Logunov, and Todorov \cite{BLT}. In
this approach, it is also assumed that the Lagrangian is written
in the form of the normal product and  there is a connection
between spin and statistics: fields with integer spin  commute
with one another and with other fields, whereas fields with
half-integer spin anticommute with one another but commute with
integer-spin fields. (\cite{BoSch}, Ch. II, \S 14)\\[-15pt]

 $\bullet$ The TCP-theorem is remarkable because a discrete
symmetry is shown to exist in theories which, to begin with, are
only assumed to be invariant under connected continuous groups...
(R.\,Jost \cite{j3}, Ch. V, \S 2)\\[-15pt]

$\bullet$ ... A very important consequence
concerns the equality \cite{lu2} of masses and total lifetimes of
particle and antiparticle, a result which is true irrespective of
the particle conjugation non-invariance of the weak decay
interactions... (\cite{mar}, Ch. 3, \S 5)\\[-5pt]

\noindent\textbf{\large 2. Theoretical and experimental status of the CPT}\\[-10pt]

The general principles of the quantum field theory that lie at the
foundation of the CPT theorem and were formulated in the mid-20th
century, connect any violation of the CPT invariance with
far-reaching changes in such fundamental concepts of the theory as
the causality principle (locality of the lagrangian) and the
relation between  spin and statistics (see, e.g., \cite{ok1}). Hence
a critical discussion of modern unconventional (and also
Lorentz-non-invariant) theories involving CPT violation and their
experimental testing are necessary elements in the progress of
physics. Further theoretical scrutiny of the current status of the
CPT and of the conditions of validity of the CPT theorem is no less
important.

Does the Lorentz invariance (LI) still hold in the theory when the
CPT symmetry breaks down, the way this occurs in models with unequal
masses of particles and antiparticles ($\bar m \neq m$)? As follows
from Greenberg's paper \cite{gre}, the answer is negative: the
general Greenberg theorem states that the interacting fields that
break the CPT symmetry inevitably break the LI as well. The CPT
invariance here is necessary but not sufficient for the LI. Theories
that break the CPT as a consequence of mass difference between
particles and antiparticles must be non-local. Then Greenberg
discusses what does the property of locality mean in quantum field
theory.

The starting points of Greenberg's work \cite{gre} are as follows.
Quantum field theory is Lorentz-covariant on the mass shell if
vacuum matrix elements of unordered products of the fields
$\phi(x_n)$ (Wightman functions $W^{(n)}$ \cite{cpt}) are covariant.
The Lorentz covariance (in fact, the Poincar\'{e} covariance) on the
mass shell is assumed from the beginning. Quantum field theory is
covariant off the mass shell if the vacuum matrix elements of
time-ordered products of fields ($\tau$ functions) are covariant.
For the LI to hold, quantum field theory must be covariant both on
and off the mass shell.

Greenberg's proof employs Jost's axiomatic approach
\footnote{As it is mentioned by Greenberg in his recent work \cite{gr3}, the Jost's approach has two advantages
over the lagrangian one: (1) it makes clear why CPT is fundamental but non of the individual C, P, and T
symmetries is; (2) it gives a simple way to calculate CPT without calculating C, P and T separately.
The goal of this pedagogical paper is to `deaxiomatize' CPT theorem considering a few items: the lagrangian CPT theorem;
representation of the real and complex Lorentz groups; vacuum matrix elements of $\tau$ functions and analytic
functions; enlargement of the domain of $\tau$-functions analyticity; the general formula for CPT; CPT for the $S$ matrix.}.
Jost formulated the fundamental theorem \cite{j1} stating that the
necessary and sufficient condition of the CPT symmetry is that the
so-called weak local commutativity holds at Jost points in the form
\begin{equation}
W^{(n)}(x_1, x_2,...,x_n) = W^{(n)}(x_n, x_{n-1},...,x_1),
\end{equation}
\vspace*{-6mm}
\begin{equation}
W^{(n)}(x_1, x_2,...,x_n) = \langle 0|\phi(x_1) \phi(x_2)\cdot
\cdot \cdot\phi(x_n)|0\rangle.
\end{equation}
Because the $\tau$ functions can be expressed in terms of the
properly arranged sum of Wightman functions, it follows that the
invariance condition is a constraint on $W^{(n)}$ given by the
relations (1), i.e., by the condition of weak local commutativity (WLC)\,
\footnote{This property was called WLC by Dyson who, as a further deduction
from the ideas of Jost and Wightman, proved the well-known theorem \cite{FJD}: a Wightman function
will be analytic and one-valued at a real set of space-time points if and only if the fields
possess a property of WLC at the same points. This statement assumes that CPT invariance is hold,
while for the case when CPT symmetry is absent a similar but more complicated statement is proved.};
this immediately implies the CPT symmetry. Consequently, any
violation of the CPT invariance in any of Wightman functions
signifies non-covariance of the corresponding $\tau$ function and,
hence, breaking of the LI of the theory. Besides, there is no reason
to deny the possibility of CPT violation in scattering and other
physical processes even if particles and antiparticles have equal
masses.\\
 In Ref. \cite{gr2} Greenberg gave a critical
analysis of an attempt to justify the model with CPT violation
caused by $\bar m \neq m$ \cite{BL} by introducing free hybrid
(`homeotic') fields that are, in the case of appropriate
normalization, linear combinations of positive- and
negative-frequency components of Dirac fields with the masses $m$
 and $-m$. It was shown that even though such free fields could
satisfy the Lorentz covariance condition on the mass shell, the
interacting hybrid fields inevitably violate the Lorentz covariance
in accordance with the Greenberg theorem \cite{gre}. The model
proposed in Ref. \cite{BL} cannot serve as an example of a theory with
CPT violation. When discussing the fundamental nature of the CPT
symmetry in quantum field theory as compared to other discrete
symmetries or their combinations, Greenberg emphasized that for LI
to result in CPT, it is necessary and sufficient to have a certain
weakened form of space-time commutativity (or anticommutativity), i.e.,
WLC \, \footnote{Note that only the normal spin--statistics relationship is possible in the
axiomatic approach to quantum field theory as discussed here since
selecting incorrect commutation relations for the field results in
the field vanishing identically (see, e.g., Ref. \cite{Schwe},  Ch. 17,
\S1).}. This remark explains why free fields with $\bar m \neq m$
can satisfy LI on the mass shell but at the same time violate the
CPT symmetry.

Summarizing his investigation of the relation of the LI of a theory
to the CPT, Greenberg \cite{gr3} again returned to the question of
what was lacking for the CPT symmetry to hold (in the presence of
LI). \textit{A free or generalized free field can be Lorentz covariant but
not obey CPT invariance if the particle and antiparticle masses are
different \cite{gre}. What fails in that case is that WLC does not
hold at Jost points... Note that although the fields in these
examples transform covariantly their time-ordered products are not
covariant. Thus if we require that time-ordered products be
covariant as part of Lorentz covariance of a theory then, as shown
in \cite{gre}, free fields that violate CPT are not covariant. See
\cite{gr2} for a detailed analysis of hybrid Dirac fields
(`homeotic' fields \cite{BL})  which can be covariant only when they
are non-interacting but even in the free case have time-ordered
products that are not covariant.} \cite{gr3}

The information given above on the general theoretical status of the
CPT problem is directly connected with the experimental tests of
CPT conservation (in elementary particle physics \footnote{See,
e.g., reviews on CPT conservation \cite{wol, bl, Anto} in Reviews of Particle
Physics, 2004, 2006, 2008.} that is presented below and, especially, in neutrino physics).\\
It should be also mentioned that the exceptional importance
of testing the CPT experimentally was first realized in connection
with the discovery of the violation of the Р, С and СР invariance
(see the review talk \cite{rub}, the review \cite{o} and talk \cite{Okn}).

The constraint that is usually quoted is the stringent upper limit
of  the CPT violation in the difference $\Delta(K^0,\overline{K^0})$
of   $K^0$ and $\overline{K^0}$ masses: $|m_{K^0} -
m_{\overline{K^0}}|/ \langle m_{K}\rangle  < 10^{-18}$ \cite{RPP}.
However, since this difference caused by the transition
$K^0\rightarrow\overline{K^0}$ is small from the very beginning,
this constraint is not exclusively characteristic of the CPT-odd
interaction: the true parameters of the CPT nonconservation in the
$K^0$-$\overline{K^0}$ system \cite{bl}  can only be bounded at the
level $10^{-3}$ -- $10^{-4}$ (see the review \cite{sh} and also \cite{tak})
 \footnote{The reason for this is that, as was pointed out in \cite{sh}, it is more logical to compare the
magnitude of $\Delta(K^0,\overline{K^0})$ not with $\langle m_{K}\rangle$ but with СР- and CPT-even  mass difference
$\Delta(K_L,K_S)$.}. The general constraints imposed by analyticity
and discrete symmetries Р, С, СР, TCP on the description of binary
systems of neutral mesons of the type $(K^0,\overline{K^0})$ were
obtained in the framework of quantum field theory in \cite{MNV}.\\
The best bound on the CPT violation in the lepton sector is defined
by the  difference between the $g$ factors of the electron
and the positron \cite{RPP}: $(g_{e^+} - g_{e^-})/\langle g_e\rangle
= (-0.5\pm2.1) \times10^{-12}$. \,\footnote{Nowadays, a factor of 22
stronger limit can be obtained \cite{BNL} from measuring the muon
spin-procession frequency in the BNL ($g-2$) experiment.}

 The current status of the CPT was also presented in recent reviews
and talks. The monograph \cite{big2}  as well as talks in a series
of meetings on CPT/Lorentz symmetry  \cite{ko}-\cite{K} and  in other conferences
\cite{ok2,NickM} discuss theoretical sources and experimental limits of CPT violation.
 \footnote{It should be noted that neither the monograph \cite{big2} nor the paper \cite{big1}
used by the authors of Ref. \cite{big2} contain a theoretical justification of the
validity of the relations for NO proposed in which $m_{\bar \nu}  \neq
m_{\nu}$. What is hiding behind this fact is in all likelihood
certain internal inconsistency in these relations.}\\
The talk \cite{ok2} dealing with the classification of the effects
of violation of all discrete symmetries also describes the relation
between  CPT invariance and hermiticity of  lagrangian of the
theory. The processes discussed are those that have not yet been
studied experimentally. These are the circular polarization of
$\gamma$ quanta in $\pi^0 \rightarrow 2\gamma$ and $\eta^0
\rightarrow 2\gamma$  decays (and also the longitudinal polarization
of muons in the decay $\eta^0 \rightarrow \mu^+\mu^-$) and circular
polarization of photons in the decay of parapositronium \footnote
{The decays listed above conserve С parity while the magnitude of
the РТ-odd effect of \textbf{sk}-type correlation between the photon
spin and momentum are controlled by the difference $\beta = g^*h -
gh^*$ where $ g$ and $h$ are the coefficients in scalar and
pseudoscalar terms of the effective lagrangian. Therefore an
experimental observation of the effect would indicate that $\beta
\neq 0$, i.e. that  CPT is violated because  lagrangian is
non-hermitian.}. It is emphasized that in contrast to the case of
$\bar m \neq m$ the above examples of CPT-odd polarizations can be
formulated in a Lorentz-invariant manner.\\
Both the invited talk at the EXA'2005 conference and the plenary
talk at the LEAP'05 conference \cite{NickM} discuss CPT from the
standpoint of violations of the basic underlying assumptions of the
CPT theorem in models of quantum gravity. The possible ways of CPT
violation are classified, and their phenomenology is described in
 terrestrial as well as astrophysical experiments. An attention is payed on
disentangling genuine quantum-gravity induced CPT violation from
`fake' violation due to ordinary matter effects, particularly when
CPT breaking of this type is connected with unitarity violations.
Further discussion of the subject see, e.g., in Ref. \cite{nickM}.

Having finished the discussion of the general status of CPT we present also information
on the violation of LI  which is based on contents of the recent reviews \cite{matt}-\cite{Lib}.

 The comprehensive up-to-date review on modern tests of Lorentz invariance \cite{matt}
 summarizes both theoretical frameworks and advances for new
precision measurements in terrestrial experiments\- and astrophysical
observations. The problems involved include the following issues:
defining Lorentz violation (QFT, modified Lorentz groups);
kinematics {\it vs} dynamics; other symmetries (CPT, SUSY, Poincar\'{e});
diffeomorphism invariance and conservation of matter stress tensors\-;
Lorentz violation and equivalence principle; Lorentz violation,
causality and stable ground  state; kinematic framework (modified
dispersion relations, \,"doubly special"\, relativity (DSR) \cite{DSR}, non-systematic dispersion);
 dynamical framework (renormalizable CPT- and Lorentz-odd operators,
non-commutative space-times, symmetry and Lorentz-odd operators,
Lorentz violation with gravity); terrestrial constraints (penning
traps, clock comparison, cavity experiments, torsion balances, neutral mesons,
Doppler shift in Li, muon experiments, higgs sector); astrophysical
constraints (time of flight (and DSR), birefringence, threshold $\gamma$ reactions,
threshold particle reactions in EFT (\v{C}erenkov effect, GKZ cutoff etc.),
threshold in DSR  etc., synchrotron radiation), neutrino physics (oscillations, \v{C}erenkov effect
\footnote{See also in the beginning of Subsection B below, where Goldstone--\v{C}erenkov effect is discussed when
 violating LI spontaneously.}); phase coherence of light; gravitational observations
(gravitational waves, cosmology, post-newtonian corrections). The principal goal of the review was to form
 a logical structure of various theoretical frameworks (with relevant types of experiments).
  While theoretical issues are present without details the list of references is voluminous (more than 280 works).\\
One should also mention the relevant problem of a varying speed of light (VSL) and its
origins (see the review \cite{JG} of recent works on VSL theories): hard LI breaking, bimetric theories,
locally Lorentz-invariant theories, color-dependent $c$-speed models, extra-dimension schemes, and field theories
 where VSL comes from vacuum polarization or CPT violation. Thereby, non-linear realization of the Lorentz group
is connected with VSL theories and DSR, in particular.\looseness=-2

 As to LI and CPT violation in the framework of Standard Model Extension (SME) \cite{AKP,K}
that is discussed below in Subsection B, all the information on experimental values (or limits) for proper parameters,
available up to January 2010, are summarized in  Ref. \cite{KosRus}. Ten comprehensive data tables
include, first of all, such sectors of the minimal SME (with Lorentz-violating operators of mass dimension four or less)
as the surrounding matter sector (protons, neutrons, and electrons) as well as the photon sector.The data are given
 also for charged leptons, neutrinos, mesons and for interaction sectors (electroweak, gluon, and gravity)
  as well as for non-minimal photon sector. Three special summary tables are composed by extracting from data tables
the maximum attained sensitivities for the surrounding matter, photon, and gravity sectors. \looseness=-1

Although the deadline on the reference list of our work is the end of 2008, it is instructive to mention here
additionally the review of June, 2009
\cite{Lib}, where the authors discuss, in particular, such themes as  the Effective Field
Theory (EFT) approach, the SME with both renormalizable and non-renormalizable operators, the naturalness
problem and higher dimension Lorentz-violating operators, as well as the DSR and the Very Special Relativity (VSR)
scheme \cite{CG} corresponding to the breaking of space isotropy. \footnote{See also the second footnote in Subsection B;
 some information on DSR and VSR is given in the middle of Subsection D.}

Let us return now to CPT violation problems in neutrino physics.
They are presented  from the standpoint of  both  standard field-theory approaches
(mostly considered in Subsections A, B, C, E, F, H, I)  and more contemporary models
\footnote{Two schemes are conventionally mentioned in connection with the mechanisms
that could produce the spontaneous CPT violation in string theories: the phenomenological \cite{AKP} and one based on
decoherence due to quantum gravity effects \cite{ell} (see also Refs. \cite{Ell,ELL} and review talks \cite{el2,mav}).}
 (reviewed in Subsections D, E (final paragraphs), F, G, J, K).\\[-5pt]

\noindent\textbf{\large\vspace*{-0.1cm}3. General consequences of  hypothetical CPT and Lorentz invariance \\
\vspace*{-0.3cm}\hspace*{0.5cm} violation in neutrino physics}\\[-5pt]

\noindent\textbf{A. CPT and neutrino physics}\\[-10pt]

The early history of studying the relation between possible CPT
violation in neutrino physics  and neutrino oscillations (NO) covers
about two decades. The first attempt was the paper by Bigi in 1982
\cite{big1}. Starting with the speculations in the literature on
possible violation of Lorentz invariance outside the Standard Model
that could be detectable in the lepton sector of the theory, Bigi
investigated not the possibility of interpreting the NO data but a
more general problem of expanding the range of phenomena and
experiments whose analysis could promise sufficient progress in
improving the sensitivity of results to CPT violation. It was
pointed out  that at least in principle the
effects of СР and CPT nonconservation could be separated: the CPT
conservation signifies the equality of probabilities $P(\nu_\alpha
\rightarrow \nu_\beta) $ = $P(\bar \nu_\beta \rightarrow \bar
\nu_\alpha $) while the СР conservation results in the equality
$P(\nu_\alpha \rightarrow \nu_\beta) $ = $P(\bar \nu_\alpha
\rightarrow \bar \nu_\beta $); the indices $\alpha, \beta $ of the
neutrino $\nu$ denote here its flavor: $\alpha, \beta = e, \mu, \tau
$. The subsequent description of NO for $m_{\bar\nu} \neq m_{ \nu}$
in the case of CPT violation was achieved by introducing a double
set of parameters without writing out the lagrangian (see also the
monograph \cite{big2}) and without introducing an explicit
definition of the masses of $\nu$ and $\bar\nu$. Obviously, this
description corresponds at the same time to the violation of the LI
(see remarks on the discussion in the first part of Section 2) with
all the consequences this implies for the theory. Therefore,
neutrino oscillations models with $m_\nu \neq m_{\bar \nu}$
mentioned in Subsection K are theoretically unfounded and in fact
incorrect.

Note also that the discussion  of the relation of the СР, Т and CPT
symmetries  with NO made it quite clear for a long time \cite{Cab}
(see also the review \cite {pak}) that if the CPT is preserved, the
effects of СР and Т violation could only occur in experiments that
would monitor an excess of neutrinos with the initial flavor. At the
same time, the CPT violation may also manifest itself (in contrast
to the СР- or Т-non-invariance) in  measuring the deficit
of initial-flavor neutrinos.

Another important note concerns the type of neutrino mass and  the lepton number conservation.
Extending the Standard Model in subsequent Subsections B and C supposes  Majorana
(Dirac) masses for the neutrinos while in the general case the situation
appears to be more complicated  \cite{bar3}. First of all, we do
not know whether neutrino processes violate the conservation of
the lepton number L and whether neutrinos are identical to their
own antiparticles. In itself, introduction of the Dirac neutrino
mass into the model keeps L conserved. Any
non-conservation of L would imply the presence of Majorana mass
terms that transform the neutrino into antineutrino. With CPT
conserved and in the presence of Majorana mass, the mass
eigenvalues are of Majorana type, i.e. the neutrino is its own
antiparticle. If the CPT is conserved but the theory has no
Majorana mass terms, then mass states are of Dirac type, L is
conserved, and the neutrinoless double beta decay is forbidden.
Authors of Ref. \cite {bar3} discuss CPT violations using a simple
example of a theory with a single neutrino $\nu$ interacting with
the electron, and its CPT-conjugate antiparticle $\bar\nu$ coupled
to the positron. They suppose that for a given  spin direction, the $\nu, \bar\nu$ mass matrix $M_{\nu}$  has the form
\begin{equation}
\left( \begin{array}{cc}
\mu + \Delta & y^* \\
    y        & \mu - \Delta
\end{array}
\right),
\end{equation}
 where the upper row corresponds to the neutrino and the lower one to the antineutrino.
For stable neutrinos the matrix $M_{\nu}$ is hermitian so that the
parameters $\mu$ and $\Delta$ (Dirac masses) are real; $\Delta\neq
0$ denotes CPT violation and $y\neq 0$ (Majorana mass) denotes
nonconservation of L. An analysis shows that the  mass
eigenstates for   $\Delta \neq 0$ cannot any more correspond to
Majorana neutrinos. However, if $y\neq 0$ then there is mixing of
$\nu$ with $\bar{\nu}$, the lepton number L is not conserved and the
neutrinoless double beta decay is allowed. A further elaboration of the presented problem
see in recent talk of Kayser \cite{Kayser}.\\[-5pt]

\noindent\textbf{\vspace*{-0.1cm}B. Extension of the Standard Model: spontaneous violation\\
\hspace*{0.6cm}of Lorentz invariance and СPТ} \\[-10pt]

Although concrete origins of LI violation in neutrino physics, which
are conventionally thought to be  related to  cosmology and gravity, are not
yet elaborated, authors of  recent work \cite{YG} has been able
to present a model-independent picture of their basic features. So,
the assumption by itself that general relativity  is hold up to the
Planck mass scale implies the spontaneousness of Lorentz violation.
Indeed, violation of LI simultaneously breaks gauge symmetry of
gravity, which can only be violated spontaneously; hence a proper Goldstone
boson must  be present. The Lorentz-violating vacuum in which the
neutrino propagates may create `static` effects in preferred ether
frame due to the vacuum expectation value ("ghost condensation"), the basis
of which is a modification of neutrino dispersion relation (similar to Eqs. (11) and (12) below in Subsection D),
and `dynamic` effects owing to coupling between the neutrino and the Goldstone boson
("gauged ghost condensation"). Static effects are described by effective hamiltonian
 $h_{ij} = \pm \mu_{ij} + a_{ij}|p| + m^2_{ij}/2|p|$, where term $\pm \mu$ is for
left/right-handed neutrinos/antineutrinos, respectively, and violates CPT ($i\neq j$ are flavor indices);
the analogous hamiltonian $h_{i=j}$ is used for dynamic Goldstone--\v{C}erenkov effects.
As it follows from kinematics, any neutrino has certain  possibility to produce single (or several) goldstone quantum
(quanta) of the corresponding mode, i.e. to change its original direction and energy.\\
 As the central point  of Ref. \cite{YG}, dynamic emission of goldstones by  neutrinos
 is specially presented below in the end of Subsection E, while
 static effects are discussed mostly  in some referred works in Subsections D, E, F, G.\\
Regardless of  paper \cite{big1} mentioned above but also in
connection with searching for new more stringent constraints on the
presence of Lorentz-non-invariant terms in the lagrangian of the
Standard Model, some perturbation-theory approaches to description
of CPT-odd effects were formulated. So as  to construct a
CPT-non-invariant generalization of the Standard Model in the
framework of an effective low-energy theory, an approach \cite{kos}
was developed for treating spontaneous CPT and LI violation in
quantum field theory and in relativistic quantum mechanics. In this
case the neutrino component of the lagrangian ${\L}$ contains only
left-handed neutrinos ${L}_{a}$:
\begin{equation}
{\L} = ~
^{1}_{\bar2}i\bar{L}_{a}\gamma^\mu\overleftrightarrow{D}_{\mu}L_a
- (a_L)_{\mu ab}\bar{L}_{a}\gamma^{\mu}L_b + ~
^{1}_{\bar2}i(c_L)_{\mu\nu
ab}\bar{L}_{a}\gamma^{\mu}\overleftrightarrow{D}^{\nu}L_b;
\end{equation}
here $\mu,\nu  = 1, 2, 3, 4,\;  a,b = e,\mu,\tau$; the first term
 is the  kinetic term,
the second and third terms correspond to LI violation, the term
with $(a_L)$ corresponds to CPT violation. When  taking  gravity
into account \footnote{For a discussion of the
problem of calculation of the NO phase in curved space-time see,
e.g., work \cite{CrGi}, part II, and also Refs. \cite{LAM,Mbo} and papers
cited therein.} this extension of the Standard Model was  investigated in Ref. \cite{KOS}.

A detailed general analysis of a possible violation of the LI
 and CPT in the neutrino sector, not using the assumption of space  isotropy
\footnote{The isotropy of space was tested
  relatively recently in measurements of the  direction independence
 of the gravitational constant $G$, in experiments with
light propagation using the theory and practical methods of wave
front inversion \cite{RAG}, as well as in experiments measuring
the amplitude $A(t)$ in the angular dependence $1 +
A(t)\mathrm{cos}\theta $ of $e^-$ emission in $\beta$ decay of
$^{90}$Sr, where $\theta$ is the angle relative to the South--North
axis. It was found that $\Delta G/G$  does not exceed the level
$10^{-10}$ (see, e.g., analysis in \cite{unn}) and that the speed
of light in air and refraction index in glass are independent
 of direction, at least to within $5\times10^{-8}$ \cite{rag}.\\
  In view of connection to the Standard Model Extension, the most recent
  analysis of possible limits of the space anisotropy $r$ is given in Ref.
 \cite{BOG} (see also references therein). Some estimates are presented
 for two types of experiments with the transverse Doppler effect considered
 in terms of conventional special relativity theory and on the basis of its
 Finslerian generalization by the author. He concludes that from first-type
 experiments (soon after the discovery of the M\"{o}ssbauer effect) aimed
 at searching for ether wind one obtains the boundary $r < 5\times 10^{-10}$;
 the present day limit could be made lower by at least three orders. As for
 future second-type measurements (with the effect of harmonic oscillation frequency
 modulation) the author expects that one would  lower $r$ down to $\sim 10^{-14}$.
 (See also \cite{Lipa} and references therein.)\\
 Another problem with anisotropy of our Universe stems from the so-called cosmological birefringence
 (see, e.g., \cite{Geng}) which can results  in a non-zero rotation polarization angle $\Delta\varphi$
  of cosmic microwave background (CMB).
 As summarized in the review talk of Ref. \cite{NI} (see also references therein and Ref. \cite{Mew}),
  CMB data of the experiments WMAP and BOOMERANG give the best current constraints
 of $\Delta\varphi \sim 100$ mrad; the Planck Surveyor will improve sensitivity upon
 $\Delta\varphi \sim 10^{-2}-10^{-3}$\; (i.e., 1--10 mrad). The updated result by April 21, 2008 \cite{Xia}
 for testing CPT with CMB is $\Delta\varphi = 2.6\pm 1.9$ deg at 68\% CL.},
was given in Refs. \cite{KM, K}.
The authors gave a clear scheme for
estimating the sensitivity of various neutrino experiments relative
to the value of three parameters -- $a_L$ and $c_L$ included into
the lagrangian (5), and the difference between the squared masses of
neutrino eigenstates $\Delta m^2$, which determines the NO. It was
shown that even in the framework of the simplest scheme (with
nonzero element $c_L$ in the case of isotropic effective hamiltonian
for the transitions $\nu_e\leftrightarrow\nu_e$, and with equal
nonzero real elements $a_L$ in the case of preferred direction along
the axis of revolution of the Earth for the transitions $\nu_e
\leftrightarrow \nu_{\mu}$ and $\nu_e \leftrightarrow \nu_{\tau}$)
it is still possible to reproduce the main features of the
experimental behavior of the probabilities of the corresponding NO.
The simplified model with two free parameters analyzed by the
authors (instead of the usual four in the case of standard
oscillations) in which $\Delta m^2 = 0$ and there is no mixing $\nu
$ with $\bar\nu$ (`bicycled model'), predicts, among other things,   a considerable
azimuthal dependence for the number of atmospheric neutrinos and a
large decrease in the half-annual variation in the flux of solar
neutrinos during some weeks before and after the equinox -- an
effect due to LI violation. The authors of \cite{KM} emphasized that
the model serves to illustrate certain key effects caused by LI
violation, and demonstrates how the presence of Lorentz
non-invariance and CPT non-conservation on the scale  $ M_{\rm Pl}$
can be identified using a certain signal in NO.

The results of analyzing the consequences of the Standard Model
extension for neutrino physics \cite{KM} were summarized in a recent
talk \cite{K} (see also Refs. \cite{K'}), which gave an exhaustive
description of the theoretical investigation of LI and CPT violation
in NO. The work is based on  conventional equations of motion for
the Dirac and Majorana neutrinos, where matrices in the spinor space
are written in a more general form:
\begin{equation}
(i\Gamma^{\nu}_{AB}\partial_{\nu} - M_{AB})\nu_B = 0,
\end{equation}
\vspace*{-10mm}
\begin{eqnarray}\Gamma^{\nu}_{AB} \equiv
\gamma^{\nu}\delta_{AB} + c^{\mu\nu}_{AB}\gamma_{\mu} +
d^{\mu\nu}_{AB}\gamma_5\gamma_{\mu} + e^{\nu}_{AB} +
if^{\nu}_{AB}\gamma_5 +
g^{\lambda\mu\nu}_{AB}\sigma_{\lambda\mu}/2, \nonumber\\
M_{AB}\equiv m_{AB} + im_{5AB}\gamma_5 + a^{\mu}_{AB}\gamma_{\mu}
+ b^{\mu}_{AB}\gamma_5\gamma_{\mu} +
H^{\mu\nu}_{AB}\sigma_{\mu\nu}/2.
\end{eqnarray}
Here, all neutrino fields (including the С-conjugate ones) are
collected into a single spinor $\nu_A$, $~A = 1, 2,..., 2N$ where
$N$ is the number of neutrino types; $\lambda,\mu,\nu =$ 1, 2, 3, 4;
$m$ and $m_5$ are the mass terms and the other coefficients in (5)--(6)
correspond to LI violation, with $a, b, e, f, g$ determining CPT
violation. If the coefficients of the type $g$ and $H$ are nonzero,
a mixing of $\nu$ with $\bar\nu$ arises. In the framework of the
scheme described here, the terms with LI violation are characterized
by dimensionless combinations of $a^\mu L,~~b^\mu L,~~ H^{\mu\nu}L$
and  $c^{\mu\nu}LE,~~ d^{\mu\nu}LE,~~ g^{\mu\nu\sigma}LE$ and can
reproduce  direction-dependent effects in oscillations.

Very recently another extensions of the Standard Model was studied \cite{Ans}, which are renormalizable
in a more general framework of "weighted power counting"\, \cite{ans}. In this approach  space and time have
different weights, the theory does not contain right-handed neutrinos, nor other extra fields, and gives Majorana mass
to the neutrinos after symmetry breaking. The author considers the simplest of minimally Lorentz-breaking
schemes  in detail; this model preserves CPT and space rotation invariance, violates LI explicitly at very high energies
and restores it at low energies.\\[-5pt]

\noindent\textbf{\vspace*{-0.1cm}C. Perturbation-theory formalism for violation of Lorentz invariance, CPT\\
\hspace*{0.6cm}and equivalence principle}\\  [-10pt]

Similarly to the approach in \cite{kos}, outlined in Subsection B, Coleman and Glashow \cite{gl6}
 developed the general formalism for introducing  CPT and/or Lorentz non-invariant
perturbative terms into the theory. The authors aimed at
 a concrete problem of testing special relativity  in highly relativistic
cosmic rays and NO \cite{gl1}-\cite{gl5} (see also \cite{gl2}).
Provided that the rotation invariance holds in a preferred reference frame (e.g., when one considers
the rest frame of the cosmic background radiation)
 the renormalizable and gauge-invariant CPT-even LI-violating additional
term in the Standard Model lagrangian results in the emergence of maximum attainable velocities (MAV) of particles
which can be not equal to photon velocity $c_{\gamma}$. Each particle type $а$ is put in correspondence
with not only the mass $m_a$ that characterizes it but also with the quantity MAV in vacuum denoted by $c_a$,
 so that $c_a \neq c_{\gamma}$. It is found that this assumption is sufficient
\cite{gl6,gl1} for oscillations to appear even with massless neutrinos
 \footnote  { \textit{Massless neutrinos cannot oscillate if special relativity is unbroken. However, they can
oscillate if different neutrinos travel at slightly different speeds {\it in vacua}.}\, \cite{gl1}.}
that are typically described in terms of the differences $\Delta c_{\nu} \equiv c_{\nu_i} - c_{\nu_j}$ and
angles of the corresponding mixing matrix for MAV eigenstates. In
the most general form, the neutrino eigenstates are characterized in
the ultrarelativistic case with given momentum $p$  by the following
sum of three hermitian 3$\times$3-matrices \cite{gl6, gl5}:
\begin{equation}
\hat{c}p + \hat{m^2}/2p + \hat{b}.
\end{equation}
 Here $\hat c$  is the matrix of MAV values for the neutrino, $\hat{m^2}$
 is the diagonal matrix of squared Majorana masses, $m^2 =
mm^\dag$, and $\hat b$  is the matrix related to the CPT
non-invariant additional term $\bar{\nu}_{\alpha}
b^{\alpha\beta}_{\mu}\gamma_{\mu}\nu_{\beta}$ in the lagrangian (the
case of timelike $b_{\mu} \sim (b,\textbf{b})$ for $\textbf{b} = 0$
is considered). The matrices $\hat c$ and $\hat{m^2}$ determine the
energy eigenstates as MAV states in the high energy limit and as
mass states in the low energy limit, respectively. In the case of
NO of two flavors, the expression for the
probability of diagonal transition on the baseline $L$ has the form
\begin{equation}
 P(\nu_\alpha \rightarrow \nu_\alpha) = 1 - \sin^{2}2\Theta
 \sin^{2}(L\Phi/4).
\end{equation}
The generalized mixing angle $\Theta $ and the phase factor $\Phi$
are written explicitly in terms of eight parameters -- three mixing
angles ($\theta _m$, $\theta _b$, and $\theta _c$), three
differences  ($\Delta m^2$, $\Delta b$, $\Delta c$) corresponding
to the matrices $\hat{m^2}$, $\hat{b}$, and $\hat{c}$, and two
complex phases ($\eta$ and $\eta^{\prime}$):
 \begin{eqnarray}
\Phi\mathrm{sin}2\Theta &=& |\Delta m^2 E^{-1}
\mathrm{sin}2\theta_m + 2e^{i\eta} \Delta b\,
\mathrm{sin}2\theta_b  + 2e^{i\eta^{\prime}} \Delta c\,
E \mathrm{sin}2\theta_c|, \nonumber\\
 \Phi\mathrm{cos}2\Theta &=& |\Delta m^2 E^{-1}
\mathrm{cos}2\theta_m  +  2e^{i\eta} \Delta b\,
\mathrm{cos}2\theta_b  +  2e^{i\eta^{\prime}}\Delta c\, E
\mathrm{cos}2\theta_c|.
\end{eqnarray}
Clearly, the type of possible violation of LI and CPT can be found
from the essentially different dependences of the terms containing
$\Delta m^2$, $\Delta b$, and $\Delta c$ on $E$.

The phase of NO when the effect is due to  the violation of
equivalence principle (EP) of general relativity (first treated in
\cite{val, gas, hal}) depends on $E$ in the same way as the term with
$\Delta c$. One would expect that a corollary of EP violation in
gravitation theories discussed in the literature
 \footnote{In Ref. \cite{HL} (see also \cite{Horv}) NO are considered  (even for mass-degenerate
neutrinos) as caused by EP violation due to the string theory effects. They contribute to macroscopic gravity and
itself caused by the massless scalar dilaton partner of the graviton \cite{DP}. In connection
to Refs. \cite{val} - \cite{HL} see also Ref. \cite{LAM} (the next to last footnote in Subsection B).}
 will also be LI- and CPT-non-conserving. As mentioned
in Ref. \cite{gl2}, the phenomenological equivalence of NO under EP or
LI violation makes it possible to find directly the constraints
on the parameters $\Delta c$ and $\theta_c$ from the range of values of |$\phi \Delta f$| and
$\sin 2\theta_G$ obtained in the former case ($\phi$ is the
dimensionless gravitation potential, $\Delta f$ characterizes the
degree of EP violation, and $\theta_G$ is the corresponding mixing
angle). In addition to referring to previous publications on the
relation of NO to the effects of EP violation
     \footnote{See also the papers \cite{mm}, \cite{ms}, and
\cite{lip} that discuss experiments carried out by the time of its
publications on solar, accelerator (including the LSND experiment)
and atmospheric neutrinos, correspondingly.},
 the paper \cite{gl2} offers an important general statement
that the experimental observation of NO in itself is insufficient
for the decisive conclusion on nonzero mass of at least one of the
neutrinos, since the oscillations may be caused by a very small
violation of LI and/or EP.\\[-5pt]

\noindent\textbf{\vspace*{-0.1cm}D. `Non-standard' violating mechanisms\\
\hspace*{0.6cm}(decoherence, modified dispersion relations)}\\ [-10pt]

`Non-standard' sources of LI or EP violation and novel NO mechanisms
are usually connected with certain properties of the vacuum on the
Planckian (or even considerably larger) scales. These aspects were
treated in  review talks \cite{MAVR, MA} \footnote{More
recently, numerous mechanisms for LI and CPT violation have been
pointed out once again in \cite{VAK}.} and lectures \cite{ma} that
offered arguments in favor of the inherent sensitivity of the NO to
CPT violation in comparison with experimental data involving other
particles. The mechanism that could explain the loss of unitarity in
quantum gravity \cite{ell, Ell} -- which would result in LI and CPT
violation in one form or another -- is so far illustrated only by a
hypothetical though visually clear picture of the manifestation of
the space-time structure (the "foam") at the quantum level; this is
caused by appearance and disappearance of black holes and large
metric fluctuations that are accompanied by formation of virtual horizons.
In these talks the author uses an idea (see, e.g., \cite{Whee}) that when a particle crosses such horizons, the
information on its state may be partly lost
 \footnote{However as it is mentioned, Hawking has stated in his talk at the GR17 (\textit{17th
Intern. Conf. on General Relativity and Gravitation, Dublin, July
2004)} that information is not lost during formation and evaporation
of black holes  because in all likelihood the true (not the apparent) horizon is never formed.
Hawking's claims are discussed in the literature -- by the author of \cite{MAVR, MA} also within the
 reference number six in his invited talk of Ref. \cite{NickM} and, in details, in one of his recent theoretical works
(see Section I of Ref. \cite{MaSar}) as well as in Hawking's paper \cite{Hawk}: \textit{It is like burning an encyclopedia. Information
is not lost, if one keeps the smoke and the ashes.} Below, the debate on the space-time foam r\^{o}le is considered
as open.}.
Correspondingly, pure state evolves into a mixed one, and it is suggested to consider the density
matrices, instead of  pure quantum-mechanical states. While, in
conventional case, the connection of $in$ and $out$ states is described by the scattering matrix $S$,
in space-time foamy situations, when unitarity may be lost,
the notion of the $S$ matrix is replaced by that of the superscattering matrix, $\not{\!S}$, introduced
by Hawking, which is a linear, but non-invertible representation between $in$ and $out$ density matrices:
$\rho_{out} = \not{\!S} \rho_{in}$ where $\not{\!S}$ is the irreversible matrix.
Thereby, $\not{\!S}\rho$ may not be defined as a product $S\rho S^\dag$.
  All that implies  the loss of the unitarity in an effective low-energy theory
and the violation of CPT -- in accordance with Wald's theorem \cite{Wald} which states
that in the above case the CPT theorem is violated,
at least in its strong form because the CPT operator is not well
defined. In this connection the author of Ref. \cite{MA} discusses the problem
 of the relation between this scheme of CPT violation
 and LI (see \cite{Milb}, \cite{GBMav} and review \cite{mavro}). A range of aspects of this problem
 \footnote{Such issues as \textit{strong form} and \textit{weak form} of CPT invariance are discussed also in
talks \cite{NickM} and plenary talk \cite{mavro}.} is considered in the author's different
publications in which he discussed other possibilities too -- e.g.
the ill-defined definition of antiparticle \cite{BeMav}, as well as
the idea of direct violation of CPT \cite{MAVR} caused by the
nonzero $\Lambda > 0$ term that accelerates the expansion of the
Universe and results in the formation of a cosmological horizon
\cite{NickM}.

In addition to a brief review of theoretical ideas concerning CPT
violation at lengths of the order of $M^{-1}_{\rm QG}$ that are
characteristic of quantum gravity, the talk \cite{MA} and lectures \cite{ma}
contained different issues of phenomenological testing
CPT in various neutrino processes including astrophysical and
cosmological manifestations. A review of a number of papers is also given in Refs. \cite{MA,ma}
of the feasibility of the above picture. Author investigates whether  it
is possible to use the available data, including NO data, for the
evaluation of parameters that characterize, first, the openness of
the system that results in quantum  decoherence
\footnote{Nonunitary evolution of a quantum system in which a
pure state is transformed into a mixed one was discussed by
Marinov \cite{Misha} who used an equation of a type similar to
(10).} (see Refs. \cite{ell,Ell,ELL})  according to the right-hand side of the Liouville equation,
\begin{equation} \dot{\rho} - i[\rho,H] = \delta H \rho,
\end{equation}
 and, second, the distortion of standard dispersion relations  (see Refs. \cite{ACE,
EV, John}) via the addition of new terms (which are represented in the
general case by a model-dependent function $F$),
\begin{equation}
E^2 = p^2 + m^2 + F(E, \vec{p}, M),
\end{equation}
(here energy-scale factor $M$ stands for $M_{\rm QG}$ or Planck mass $M_{\rm Pl}$)
that result in  CPT and/or LI violation.\\
In the conventional approach when the preferred  frame is at rest relative to the cosmic microwave background radiation,
to change LI minimally with keeping energy-momentum conservation it must be only the boost invariance broken but the
rotation symmetry is preserved while, in the reverse order case, the rotation invariance breaking entails the boost one
too. Then, as assumed in the most of QG models, modified dispersion relations (MDR) come to the form:
\begin{equation}
E^2 = p^2 + m^2_{\alpha} + f^{(1)}_{\alpha}p^2|p|/M + f^{(2)}_{\alpha}p^2(|p|/M)^2 + f^{(3)}_{\alpha}p^2(|p|/M)^3 + ...~ ;
\end{equation}
 the parameters $f^{(u)}_{\alpha}$ ($u = 1, 2, 3...$) are dimensionless and are labeled in accordance to particle species.

We notice here the theoretical problem of the presence in quantum field
theory and in Standard Model, particularly, self-energy contributions to MDR in Eq. (11). As it has been shown  in Ref.
\cite{Coll} (see also Refs. \cite{Burg,Myer} and the literature
discussing these papers), the resulting contribution to MDR even
with counterterms and renormalization procedure cannot preserve LI
from violation at percent level without any suppression, unless the
bare couplings of all the particles involved are strongly fine
tuned. Hence the matter is also to search for mechanisms to maintain LI but not just to lower its breaking limits.\\
As for the quantum decoherence (QD), the general case of a phenomenological description of NO
 with two flavors treated as an open system was analyzed in detail in Ref. \cite{ben}. Dissipation effects in the
right-hand side of (10) were treated in the approximation in which
quantum gravity results in linear decoherence (with linear dependence on density matrix)
\footnote{This linear approximation may not comply with the complete theory \cite{EMN} (see also Ref. \cite{GBMav}).};
 they are simply parametrized by six real variables. These quantities are related via a number of
inequalities that correspond to the property of `total positivity'
required to ensure that the density matrix, which describes the
states of the extended system that includes not only neutrinos but
also their microenvironments with a characteristic scale length,
is positive. Three additional parameters are then introduced into
the effective hamiltonian; they correspond to the interaction with
the surrounding part of the system (for simplification, the authors kept only one of two which is additive to the
conventional parameter $\Delta m^2/2E$).

The authors of Ref. \cite{EMN} obtained and analyzed formulas for NO
probability in the case of general dependence of decoherence effects
on all parameters; they emphasized that these effects manifest
themselves even with massless neutrinos and depend on the СР-odd
phase which is present in the mixing matrix for the Majorana
neutrino. In principle, this feature may serve to distinguish this
case from that of the Dirac neutrino.\\
Later,  a study of simplified model for flavor oscillation was carried out \cite{MSar}, coming from Liouville
decoherence and emphasizing attention to the cosmological constant (dark energy) as an origin of exponential
quantum suppressing of low-energy observables. Invited talk  \cite{mavrS} discusses the r\^{o}le of space-time
foam already as the small possible contribution to $\Delta m^2$ in NO and speculates on
the connection of QD with the dark energy, i.e., on involving QG-foam effects in the cosmological $\Lambda$-term origin.
Then, elaborated ideas and methods of QG-induced QD in the string theory framework with underlying phenomenology
was presented in invited talk \cite{MavrS}.\footnote{Contemporary methods and models for studying decoherence
in particle physics, in general setting and, particularly, with its role in qualitatively new phenomena which imply
discrete-symmetry violations, are described in review \cite{sark}.}
The decoherence  evolution is considered with its characteristic
feature of exponential time-dependent damping in NO probability both linear in $t$, as in the conventional case, and
quadratic in $t$, as in cases of stochastically (random) fluctuating space-time foam \cite{GBMaV,MaSar}.

Besides considering several topics, above-quoted Ref. \cite{MA} discussed also  non-linearly modified
Lorentz transformations: both in connection with unitary
non-equivalence of Fock's flavor and mass spaces in the NO description in quantum field theory \cite{BMP}
 \footnote{The low-energy limit of QG with $\Lambda >0$ which must be invariant under deformed Poincar\'{e}
 symmetry \cite{ASS} is discussed in the end of Ref. \cite{mavro}.},
 and in view of the natural requirement of invariant definition of the scale of Planck length/energy
\cite{Amel, MaSm}. Here, in a speculative scheme  of "doubly special
relativity"\,(DSR) \cite{DSR} \footnote{In the next to last paragraph of Section 2 see also some
issues of the DSR within the problems of review \cite{matt}.},
 the Lorentz group acts non-linearly on physical quantities, and new choice of group action leads
 to a new invariant energy scale (usually, Planck mass) as well as the invariant velocity $c$.
In DSR models there is a dependence of the speed of light on the wavelength, and LI violation is really
 only `apparent'\, effect when the usual {\it linear} Lorentz group action is violated. \\
While early DSR-type studies used broken LI, the most popular DSR schemes have `deformed' LI, with no actual
`loss of symmetry' when in presence of a second observer-independent length scale $\lambda_{\rm DSR}= M^{-1}_{\rm Pl}$
all inertial frames remain equivalent, but modified Lorentz transformations appear as
the transformation laws between the frames.
In particular, MDR of the form $0 = 2[\cosh(\lambda E)- \cosh(\lambda m)]- p^2 \exp(\lambda E) \simeq
E^2 - p^2 - m^2 - \lambda Ep^2$ (here $\lambda M_{\rm {Pl}}\equiv f^{(1)}$ in (12)) can hold for all inertial frame
if a deformation of the  boost transformations is $\lambda$-dependent \cite{GAKG}.\\
Basing on exact form which is found for energy-momentum conservation law that is characteristic for deformed LI only,
in contrast to the broken LI case when the usual form is valid, the authors managed to distinguish both scenarios by
the sign of parameter $\lambda$ \cite{GAKG}: DSR requires $\lambda >0$ as well as has typically small threshold anomalies
 and photon stability, the very properties that is absent in broken LI scheme where the possibility of positive $\lambda$
 is not acceptable and is already excluded experimentally.

Although the DSR as a generalization of special relativity, but solely in momentum space,
has the same number of generators (i.e., ten ones), it is not clear how to relate momentum to position
 while the notion of DSR space-time is still not developed; therefore now DSR is only a kinematic scheme \cite{Lib}.

 As for the well-known problem whether  flavor eigenstates or mass eigenstates emerge as real objects in a QFT-treatment
 of NO, that mentioned above in view of Ref. \cite{BMP}, the authors' further studies of the existence of a Hilbert space
 for the flavor states showed that mixed neutrinos have proper MDR and that the corresponding non-linear realization
 of the Lorentz algebra is of the DSR-type \cite{MaSm}.

Besides perturbative schemes of Kosteleck\'{y} et al. and of Coleman and Glashow in
Subsections B and C, respectively, another, more drastic approach to the problem of possible
failure of LI was suggested by the authors of Ref. \cite{CG}. In the Very Special Relativity (VSR)
they substitute for the Poincar\'{e} group (as the exact symmetry of nature) by one of its certain subgroups
which include space-time translations along with at least a 2-parameter subgroup of the Lorentz group (LG).
Their subsequent work \cite{CoG} where VSR contains only a 4-parameter subgroup of LG, supposes
a non-standard origin of lepton-number conserving neutrino masses without need for Yukawa couplings and
see-saw models but with  $2\beta0\nu$ processes forbidden as well as tritium $\beta$-decay end-point spectrum
different from standard \footnote{As for the tritium $\beta$-decay end-point spectrum in the VSR framework see discussion
after the middle part of Subsection G.}.

In contrast to the analysis of renormalizable Lorentz non-invariant
terms in Ref. \cite{kos} and Refs. \cite{gl6, gl1} that is described in
Subsection C, the general discussion in Ref. \cite{Brus} of the possible LI violation on
Planck scales that would affect NO was focused on studying
nonrenormalizable effects that result in the energy
dependence of oscillation length of the type $L_{\rm osc} \propto
E^{-n}$  with   $n = 2$. The dependence with $n \neq -1$ is essential evidence of LI and CPT violation \cite{Yas}.
In the literature, there are the cases of this group with $n = 0$ \cite{DeS,gl1,LAM}
\footnote{Cases of energy-independent NO at $m_\nu = 0$ are also
treated in Ref. \cite{Klin}; hire,  by analogy with solid state physics, the LI and
CPT violations are introduced in the fermion vacuum of quantum field
theory.}, $n = 1$  \cite{gas,gl6}, and also $n = 2$
\cite{Eich,Alf,Lam} \footnote{$L_{\rm osc}$ is proportional to
$M^2_{\rm QG}/E^2$  if it is assumed that LI/EP is violated as a result
of unequal values of MAV due to recoil effects in neutrino
scattering by virtual $D$ branes \cite{EV}.} and $n = -3$ \cite{Adu}
\footnote{$L_{\rm osc}$ is proportional to $E^{-3}$ in the case
considered of $m$ = 0 in a $q$-deformed non-commutative theory
\cite{Chen}. EP violation in effective Schwarzschild geometry
modified by the hypothetical presence of the maximum acceleration
$\mathcal{A}_m = 2mc^3/\hbar $ in the chosen  gravity model
corresponds to $L_{\rm osc} \propto \Delta m^2/E^3$ \cite {Boz} (see
also Ref. \cite {Mbo}).}. Note that all possible terms in the effective
action that are renormalizable and invariant under rotations
correspond to $n = 0,\, \pm1$.

The consequences of EP violation in noninertial reference frames
were considered in Ref. \cite{LAM} with $n = 0$; if $\Delta m^2 = 0$
and linear acceleration is zero, $L^{-1}_{\rm osc} \propto
\omega\cos\beta$ where $\omega$ is the angular velocity of the
system (in the case of the Earth, $\omega \sim 7\times 10^{-5}$
rad/s) and $\beta$ is the angle between the rotation axis and the
momentum of the neutrino. It is emphasized the fact  (already
discussed in the literature) that the choice of metric affects the
estimates of EP violation from NO data.

The authors of papers \cite{Adu}, in which the dependence on energy
corresponds to $n = -3$, $L_{\rm osc} \propto E^3/(\Delta m^2)^2$, started with a speculation
that the inertial and gravitational masses, $m_i, m_g,$ are two independent objects and the flavor oscillations carry
the fluctuations $\Delta E \Delta t \sim\hbar$ in a coherent manner: the inherent
energy uncertainty for flavor states is related to the inverse of time period of NO.
This general statement leads to quantum-induced violation of EP: $m_g = (1 + f)m_i$, where
$f $ is non-zero for systems with no classical counterpart. Modification of standard commutation
 relation  on Planck scales is also discussed in the last paper of Ref. \cite{Adu}:
$[\textbf{x, p}] = i\hbar(1 + L^2_{\rm
Pl}\textbf{p}^2/\hbar^2)$ where $ L_{\rm Pl} = \sqrt{\hbar G/c^3}
\sim 10^{-33}~$cm.

One more origin of the spontaneous  violation of LI invariance can arise
in non-commutative field theories  (see, e. g., \cite{Nekr}). In such a framework \cite{Ari} and
in the simple case, equal-time anticommutation relations have the
form $\{\psi^i (\textbf{x}), \psi^{j\dag} (\textbf{y})\}=
\mathcal{A}^{ij}\delta^{(3)}(\textbf{x}-\textbf{y})$ ($i,j$ stands for flavor indexes)
where $\mathcal{A}^{ij} = (^{1~~ \alpha}_{\alpha^* ~1})$ is a constant matrix,
$\alpha$  is a deformation parameter in flavor space. In treating this
deformation as the some kind of the extension of Standard Model
\cite{K'}, the massless-type MDR appear with non-trivial scale factor (so that $\nu$ and $\bar\nu$
of different flavors are degenerate in energy). This implies  LI violation, leads to mass-independent
 energy difference for flavor states and to $L_{\rm osc} \propto |\alpha|EL$, while CPT
symmetry remains intact.\\[-5pt]

\noindent\textbf{\large\vspace*{-0.1cm}4. Experimental and observational consequences of CPT and Lorentz \\
\hspace*{0.7cm}invariance violation in neutrino physics}\\[-10pt]

This Section contains the information on works dealing with
those specific models of CPT, LI and EP violation in various
neutrino processes where, as a rule, flavors are changed and for which the estimates
of parameters that characterize the appropriate violation were
obtained by comparing with measured data. The information on early works which discuss these subjects  see in Section 5
of Ref. \cite{GGMM}.\\[-5pt]

\noindent\textbf{\vspace*{-0.1cm}E. Perturbative violation of Lorentz invariance and CPT}\\[-10pt]

In this Subsection  we review papers that consider constraints on
the parameters of perturbative violation of LI and CPT  (with  EP not violated)
\footnote{See also the end  of Subsection I.}; these parameters are
predicted or expected on the basis of analyzing NO manifestations.

A comparison of the expressions (6)-(7) in Subsections B and C with the neutrino data of
the 1990s showed that Lorentz non-invariant terms are found to be
too small and do not significantly affect the interpretation of the
available NO results (except for CPT-odd effects at very long
baselines). At the same time, further investigation of oscillations
of solar neutrinos and accelerator neutrinos at $E\sim$ TeV and
baseline $L\sim10^3$ km  may detect LI violation when $\Delta
c\sim10^{-25}$ \cite{gl6, gl5}. A recent analysis in Ref. \cite{ShG}
showed that a more stringent  constraint than earlier ones may be
obtained from the Super-Kamiokande (S-K) and MACRO experiments  for
atmospheric neutrinos at $E\sim100$ GeV and $L\sim10^4$ km: $\Delta
c<10^{-25}$.

The authors of Ref. \cite{bar} treated the cases of manifestation of
CPT-odd effects  caused by the interference of terms with $\Delta m^2$
and   $\Delta b$ in (9) when resonant amplification of NO
amplitude becomes possible at $\mathrm{sin}^2 2\Theta = 1$, by
analogy to the well-known MSW resonance  when neutrinos
pass through a sufficiently dense medium
 \footnote{The phenomenon of the resonant flavor change of   the
MSW transition type was reported earlier in Refs. \cite{val, gas, hal}
and in other papers on EP violation (see, e.g., the references in Ref.
\cite{gl2}).}.
 For instance, the resonance occurs in a medium
 with number density of electrons $N_e$ for a simplified situation
  of $\theta_m = \theta_b \equiv \theta$ and $\eta = 0$ when   the denominator of the generalized mixing angle
$\Theta$,
\begin{equation}
\mathrm{tan}2\Theta = \frac{(\Delta m^2 + 2E\Delta b) \mathrm{sin}2\theta}
{(\Delta m^2 + 2E\Delta b)\mathrm{cos}2\theta - 2\surd2G_{F}EN_e)},
\end{equation}
vanishes for flavor index $\alpha = e$  in (8); here $G_{F}$
is the Fermi constant.  The Ref. \cite{bar} argues that  it is possible,
in principle, to achieve estimates as low as  $\Delta b \sim
5\times10^{-23}~$ GeV, when analyzing the CPT violation in
atmospheric neutrinos. And in neutrino factories, CPT violations could be detectable at the $3\sigma$ level
  for $\Delta b \approx (1 - 3)\times 10^{-23}~$ GeV depending on the baseline length $L$ (at  29 GeV
 energy of stored muons).\\
At the same time, new analyses of the data  on the absence of
$\nu_{e,\mu} \rightarrow \nu_{\tau}$ oscillations in the latest
accelerator short-baseline experiments CHORUS and NOMAD are
expected to furnish the limiting values at the levels of $\Delta b<
10^{-18}~$ GeV and $\Delta c < 10^{-20}$  \cite{KM}.\\
As for the value of $\Delta b$, a joint analysis \cite{bah} of the
data on the solar neutrino and the expected sensitivity of the
KamLAND reactor experiment gives the upper bound  at the level $10^{-20} - 10^{-21}$ GeV. The result of mid-2004
global fitting \cite{BGP} of solar and reactor data points to $\Delta b < 0.6
(1.5)\times 10^{-20} $ GeV at $1\sigma~(3\sigma)$, respectively.

More stringent restriction can be derived by taking into
account that the neutrino and the charged leptons sectors of the
theory are closely connected. So, the estimate $\tilde{b}_i<
10^{-17}~$ eV for spatial components of the quantity $\tilde{b}$
(a conventionally chosen additive combination of $b$, and
coefficients $d, H$, and $g$) in the sector of left-handed neutrinos
was obtained for models with heavy right-handed Majorana neutrinos
\cite{moc} on the basis of the available very strict limit on the
axial term $\bar e b_\mu \gamma_\mu \gamma_5 e$ defined by the
relation $|\tilde{b}_i(electron)| \lesssim 10^{-19}$ eV. This
estimate for $\tilde{b}_i$ is weakened by four orders of magnitude
($\tilde{b}_i \sim 10^{-4}\tilde{b}_0$) by taking into account the
motion of the solar system relative to the Galactic halo and that of
the Earth around the Sun, so that a selection of a reference frame
for $\tilde{b}_{\mu}$ brings the obtained constraint down to the
level $10^{-13}~$ eV -- still much more stringent than is
anticipated for direct neutrino experiments \footnote{The upper
limit shown above  for $|\tilde{b}_i(electron)|$ was obtained
in a precision experiment with torsion balance in which the probe
body possessed certain residual magnetization caused by the spin
dipole moment (due to the polarization of electrons). Later on, this
experiment reduced   the limit down  to the level $10^{-20}$
eV \cite{h}. Hence, CPT-violating constraint appears to be harder
by a factor of ten.}.

To make the manifestations of the possible LI and CPT violation
in oscillations accessible for realistic observations, the
neutrino sector should be `shielded' \cite{cho}  from the sector
of charged leptons. The authors of Ref. \cite{cho} connected the
implementation of this idea with a unique operator
$h^{\mu\nu}_{\alpha\beta}({\overline{\nu^C_L})_\alpha}\sigma_{\mu\nu}
 (\nu_L)_{\beta}$ that emerges
in the light left-handed neutrino sector via a see-saw-type
mechanism through introduction of the appropriate LI violation for
the heavy Majorana neutrino characterized by the constants
$H_{\alpha\beta}$. This approach results in non-conservation of the
lepton number L ($\Delta $L = 2) while LI violation (with CPT
conserved) valid for conventional neutrinos does not cover charged
leptons via the radiative corrections in all orders of
perturbation theory. The appropriate oscillations length is found
to be independent of energy (as it is in the case of flavor
transitions due to  magnetic moment of the neutrino) and
is dictated only by the constants $H_{\alpha\beta}$. A comparison
of this approach with data (or estimates) provided the authors with the following constraints:
$H_{\mu\tau} \lesssim 10^{-20}~$ GeV (for atmospheric
$\nu_{\mu}$), $H_{\mu\beta} \lesssim 10^{-22}~$ GeV (for
accelerator $\nu_{\mu}$ on a long baseline), $H_{\mu\beta} \lesssim10^{-23}~$ GeV (for
 neutrino factories), $H_{e\beta} \lesssim10^{-19}~$ GeV (for
$\nu_{e}$ of the reactors CHOOZ and Palo Verde); the results of
the KamLAND experiment with reactor neutrinos are described in
this case at $H_{e\beta} \lesssim  7.2 \times 10^{-22}~$ GeV.

Finally, the ratios of the expected numbers of $\nu_{\mu}$ and
$\bar\nu_{\mu}$ events were estimated  \cite{dat} in connection
with  new multi-kiloton magnetized iron
calorimeter projects for studying atmospheric neutrinos in laboratories at
Gran Sasso (Italy) and INO (India); the authors  compared the
results  with predictions found in the CPT and LI violation scheme. The
dependence of this ratio on $L$, $L/E$ and  $LE$ obtained for a
number of values of $\Delta b$
confirms the possibility of detecting these violations for $\Delta b > 3\times
10^{-23}$ GeV. 
 These estimates are more stringent than was predicted for  future neutrino factory projects.

Additional information on constraints of the quantities $\Delta c$ and
$\Delta b$ in (9) that comes from NO data was presented in Ref.
\cite{pak}. Similar results on constraining the parameters of
 possible violation of LI and CPT that come from experiments
with atomic systems and muons were reported in the review talks of Ref.
\cite{blu}. Constraints on LI violation parameters on Planck
scales are also obtained by analyzing the high-energy parts of the
cosmic rays spectrum; these constraints indicate that there is no effect of
the \v{C}erenkov radiation in vacuum for $p$, $e$, $\mu$ and $\nu$ \cite{JLM}.

As for using astrophysical and cosmological neutrinos to test LI violation via  their Goldstone--\v{C}erenkov
emission \,\footnote{In the framework of SME the goldstone emission mechanism is considered as secondary in the context
of testing LI in NO and is disregarded (see, e.g., Ref. \cite{KKT}).}
(see Ref. \cite{YG} in the beginning of Subsection B), the authors evaluate three quantities -- the emission rate,
the neutrino energy loss rate, and the average deflection angle for single emission event. Comparing this values to
SN1987A $\nu$-events data and CMB information on the energy accumulated in cosmological $\nu$ results in the following
set of LI-violating bounds  \cite{YG}:\\[3pt]
\vspace*{2mm} $ ~~~~\mu_{SN}\lesssim 10^{-11}$GeV$(\frac{M}{10~\mathrm{MeV}})^{3/2}(\frac{0.1~\mathrm{eV}}{m}),
~~~a_{SN}\lesssim 10^{-17}(\frac{M}{10~\mathrm{MeV}})^{3/2}$;\\
\vspace*{1mm} $ ~~~\mu_{CMB}\lesssim 10^{-22}$GeV$(\frac{M}{1~\mathrm{eV}})^{2}(\frac{0.1~\mathrm{eV}}{m}),
~~~~~a_{CMB}\lesssim 10^{-11}(\frac{M}{1~\mathrm{MeV}})^{2}$;\\
\vspace*{1mm}$~~~~\mu_{SN}\lesssim 10^{-15}$GeV$(\frac{M}{10\,\mathrm{MeV}})(\frac{10^{-3}
~\mathrm{eV}}{g})^{3/2},~~~a_{SN}\lesssim 10^{-15}(\frac{M}{10\,\mathrm{MeV}})(\frac{10^{-3}~\mathrm{eV}}{g})^{1/2}$.\\
Here, the first SN and CMB bounds are due to "ghost condensation"\,(`static' effects) while lower SN bounds are due to
"gauged ghost condensation"\,(`dynamic' effect),
$M$ is the scale of spontaneous Lorentz violation, $m$ is the neutrino mass, $g$ is the gauge coupling,
parameters $\mu$ and $a$ correspond to dimension five and eight operators, respectively.\\[-5pt]

\noindent\textbf{\vspace*{-0.1cm}F. Violation of Lorentz invariance and equivalence principle\\
\hspace*{0.6cm}in terrestrial and cosmic neutrino physics}\\[-10pt]

In the review talk \cite{leu} that outlined the fundamentals of EP
 and LI violation in NO when neutrinos interact with the background
 gravitational field, the corresponding results of analysis
 of solar and atmospheric neutrino data available at the end of 1990s,
 as well as references to earlier works were presented.
 \footnote{In addition, see the talk \cite{Yas}, paper \cite{mi} and
 references therein, paper \cite{LAM} mentioned earlier
(in the second part of Subsection D),  and also papers \cite{cas} where
the gravitational interaction with the neutrino takes into account,
besides the potential $\phi$,  next-order terms in the
post-newtonian approach, that describe new   anisotropy effects.
Violations of Einstein's EP  are considered also in Ref. \cite{NI} (see last footnote in Subsection B),
 in connection with the cosmological birefringence
(sizable rotation polarization angle via, in particular, the number density difference between neutrino and antineutrino).}
 The best constraints (the safest estimates) on the parameters $\Delta c$
 and  |$\phi\Delta f$| for atmospheric neutrinos were \cite{fog} $6\times10^{-24}$
 and $3\times10^{-24}$ at 90$\%~$CL, respectively,
  regardless of the values of  mixing angles. The result for solar
 neutrinos was found to be at a similar level but was affected by
the choice of assumptions. A detailed analysis of the available at
that time atmospheric neutrino data with arbitrary values of the
parameter $n$ in   the energy dependence of the oscillation length,
$L^{-1}_{\rm osc} \propto E^n$, resulted in the constraint
\cite{fog} $n = -0.9\pm0.4$ at 90\% CL ($n = -1$ corresponds to
ordinary oscillations of massive neutrinos). Still earlier results
of experiments with atmospheric neutrinos failed to provide an
opportunity to exclude any of the existing scenarios of EP violation
in the spin-$J$ field exchange \cite{foo}: scalar with $J$  = 0 and
$n = -1$ (dilaton), vector with $J$ = 1 and $n$ = 0 (torsion in the
Einstein--Cartan theory), tensor with $J$ = 2 and $n$ = +1
(graviton).

The authors of Ref.  \cite{gag} give a description of global fitting
 of all solar neutrino data obtained before the publication of the
 results of the experiment SNO-2002 with solar neutrinos, and also
 provided information on previous interpretations of oscillation data,
including those based on EP violation \footnote{See also talks
\cite{GAG} and papers \cite{berg, Maj}.}. The purpose of this work
was to obtain a numerical comparison of  possibilities of explaining
experimental results in terms of different flavor-changing
mechanisms for $\nu_e$. It was shown that in addition to the known
large-mixing LMA(MSW) solution, oscillations can be explained at
the same confidence level ($\gtrsim 60\%$) due to several mechanisms,
in particular,  by EP violation, by
neutrino flavor changing through interaction of its magnetic
moment with the external magnetic field, and by nonstandard
neutrino interactions (NSNI) parametrized by two constants, one of
which characterizes the contribution of flavor-changing
interactions while the second determines the ordinary
neutrino--medium interaction and plays a role similar to that of
$\Delta m^2$ at the MSW resonance. It is also emphasized that
experimental data do not warrant obtaining stringent constraints
for the existence of solutions based on NSNI or on EP violation
 \footnote{Many papers appeared recently in the literature, which
interpret NO data in terms of NSNI mechanisms. Their
phenomenological manifestations are typically characterized by that
the effect is independent on neutrino energy, which inhere also in
the contribution of the parameter $\Delta b$ in (9). It is seen from
the discussion in Subsection C that this is also the case for the
scalar version of EP violation. Hence, information on the limiting
values of  $\Delta b$ and |$\phi \Delta f$| is likely to be
extractable also from the data that yielded constraints on the
parameters of NSNI.}. The best description of solar NO reported in Ref.
\cite{gag} corresponds, in the case when they are caused by EP
violation, to |$\phi \Delta f$| $\simeq 1.6 \times10^{-24}~$ and
$\sin 2\theta_G$ = 1 (vacuum-type solution); the MSW-type resonance
solution requires \cite{pan} the values of |$\phi \Delta f$| that
would be incompatible with the CCFR data.\\
The last analysis in Ref. \cite{Vald} of EP violation (within a simple model of the gravity going on
physical mass basis) for solar $\bar\nu_e$ data from experiments Homestake, Sage, Gallex/GNO, S-K,
and SNO (including reactor KamLAND results) gives the values for NO
parameters near standard mass-flavor
 MSW solution and the conclusion, that the superior limit is $|\phi\Delta f| \leq
1.3\times10^{-20}\, (3\sigma)$ and the effect might  take place for reactor antineutrinos.

A fitting of atmospheric neutrino data within EP violation
 or in the presence of NSNI provides very poor results (see talks
\cite{lus} and references therein); no interpretation of these data
 on the basis of pure NSNI mechanism is acceptable for $99\%$ CL \cite{For},
mostly due to the independence of this mechanism on energy. The
subsequent  fitting \cite{GGM} of atmospheric S-K data and the К2К
experiment results showed that resorting to LI and CPT violation as
an additional mechanism of NO hardly affects standard parameters.
 The  restrictions (giving eightfold improvement on the
 results of Ref. \cite{fog} (see the beginning of Subsection F) obtained at  90\% CL
in the $\nu_\mu-\nu_\tau$ sector are the following  \cite{GGM}: $|\Delta c| \leq 8.1 \times
10^{-25}$, $|\phi\Delta f| \leq 4.0 \times 10^{-25}$, $|\Delta b|
\leq 3.2 \times 10^{-23}$ GeV, and $|\Delta \delta_0| \leq 4.0 \times
10^{-23}$ GeV -- the last constraint is for NSNI with
vector-type torsion field ($J$ = 1) via CPT-even effects for $n$ = 0. The corresponding 3$\sigma$ limits were
 also obtained and found to be greater by a factor of 1.5 -- 2.
\footnote{At maximal mixing the results given are much stronger,
e.g., $|\Delta c| \leq 2.0 \times 10^{-27}$; see also the
$\nu_{atm}$ + K2K limits and  sensitivities of the AMANDA-II,
IceCube and MACRO detectors in Proc. of the first workshop EPNT06
\cite{EPNT}.} The NSNI three-flavor interpretation of experimental
results with solar and atmospheric neutrinos \cite{GUZ} supports
solar data description of Ref. \cite{gag}  (in the  case  when it is
made  more complicated due to mutual influences of these two
sectors) while the inclusion of the KamLAND data disfavors NSNI
hypothesis. Basing on upper limits of CPT/LI-violating coefficient
differences $\Delta b$, which were odtained in Refs. \cite{bar,dat}
(see Subsection E) and in Ref. \cite{GGM}, authors of a later
three-flavor
 analytic treatment \cite{dir} calculated bounds of the same order but for quite different exact combinations
of the matrix $\hat{b}$  elements (see Eq. (7)) for the FNAL two-detector experiment NO$\nu$A at $E_{\nu_{\mu}} < 5$ GeV
 and for neutrino factories  at $E_{\nu_e} > 15$ GeV, correspondingly.\\
The posterior work \cite{GHH} has shown, that these limits could be improved by
over two orders of magnitude at high energy and high statistics with IceCube detector.\\
Using a subsample of 300 upward-throughgoing muon events in the MACRO detector,  upper limits
were established \cite{MACRO} on LI- and EP-violating differences $\Delta c$ and $\Delta f$.
As the subdominant origins of atmospheric NO via $\nu_{\mu}\rightarrow \nu_{\tau}$ transition, these differences
satisfy the following bounds at 90\% CL: $|\Delta c| <2.5 \times10^{-26}$ at maximal mixing,
$|\Delta c| <3 \times10^{-25}$ at marginalization with respect to all the other parameters; as for EP violation,
$|\phi\Delta f|\Leftrightarrow |\Delta c|/2$.

It can be expected also \cite{Datt}, that for a wide range of parameter values  the estimates of EP violation
there will be obtained in muon storage rings by recording changes of neutrino flavor by ordinary NO manifestations.\\
Some more formerly in the Ref. \cite{K-K}, a scheme was suggested in order to calculate $2\beta0\nu$ decay rate
with LI/EP violation. The main conclusions, when comparing the NO and the $2\beta0\nu$ as probes of the violation
in such scheme, were the following: (1) $2\beta0\nu$ gives the most stringent bound on LI violation while NO
cannot limit it at mixing angle $\theta_c \rightarrow 0$; (2) the dominant contribution to Lorentz violation
comes from momentum-dependent $\Delta c$ term with the effective neutrino mass $\langle m\rangle$; (3) using
$2\beta0\nu$ data of the Heidelberg--Moscow experiment and known estimates for $\langle m\rangle$, the bound
$\Delta c \lesssim 10^{-16}-10^{-18} ~(\theta_c = \theta_m =0)$ was obtained. (Similar independent calculation
 scheme for EP violation, as it was shown, points out that the LI bound can be directly translated
to analogous EP bound for $|\phi\Delta f|$ when gravitational mixing vanishes while this region is
 constrained only by the $2\beta0\nu$.

A detailed study of possible constraints on the EP- and LI-violating parameters in neutrino factories \cite {LW}
 led to a conclusion that measuring the T-odd probability difference
$P(\nu_{\alpha}\rightarrow \nu_{\beta}) - P(\nu_{\beta}\rightarrow
\nu_{\alpha})$ provides the most sensitive evaluation of this
violation. For  $(\nu_e,\nu_\mu)$ and $(\nu_e,\nu_\tau)$  sectors,
the limiting value |$\phi \Delta f| \lesssim 10^{-26}$ can be achieved
\cite{LW} with a suitable baseline of several thousand kilometers, that
is comparable to the maximum constraint in the $(\nu_\mu,\nu_\tau)$   sector obtained in Ref. \cite{fog} for
atmospheric neutrinos ($\nu_{\rm atm}$) using the S-K results.

In paper \cite{GNT} a brief review of the literature is given on
consequences of EP violation in various neutrino processes,
including neutrino astrophysics and primary nucleosynthesis
\footnote{See also Ref. \cite{CaLa} about NO in wormhole-type
objects.}.
 It was shown that the low probability of  $\bar\nu_e \rightarrow
\bar\nu_ {\mu,\tau}$ transitions coming from  supernova SN1987A data points to a very strong
constraint on the corresponding parameters: for massless or mass-degenerate neutrinos
 |$\Delta f|\lesssim \mathcal{O}(10^{-31})$
\footnote{Evaluations of EP violation expressed in terms of constraints on the parameter |$\phi \Delta f|$
typically assume that the quantity $\phi$ = const $~\sim 3\times 10^{-5}$ is determined by the mass of the local galaxy
supercluster.}
 and tan$^{2}\theta_G\ll 10^{-4} $. However, these constraints
become invalid or weakened if the effect due to the mass is dominant. As it should be
 noted in this case, an analysis of consequences of EP
violation in NO, which may be detected in future observations of
super-high-energy neutrinos arriving from cosmologically remote
active galactic nuclei, is likely to yield an even stronger
constraint -- at a level |$\Delta f|\sim 10^{-41}$ \cite{MiSm}.

EP violation may be directly related to the formation of neutron
stars, namely to the fact that pulsars at the moment of birth
acquire considerable peculiar velocities. First evaluations of EP
violation in resonant flavor transitions (for the maximum
efficiency case of $J =$ 2) that can sustain required velocities
in the anisotropic ejection of neutrinos from a presupernova
(provided there is a magnetic field $~>10^{15}$ Gauss) yielded the
value  $|\Delta f| \simeq 10^{-10} - 10^{-9}$ \cite{hor}. The
translational and rotational motion of pulsars caused by
directionality of neutrino ejection can be interpreted even at
zero magnetic field \cite{bark} if resonance transitions are
assumed to be caused by above mentioned anisotropy effects \cite{cas} in
post-newtonian approach to gravitational neutrino interactions.

Finally, Ref. \cite{den}  attracted attention to the
importance of simultaneous neutrino and optical monitoring of
type-II presupernovas. The data on the time of recording and the
characteristics of both signals, as well as the observation of the
frequency difference in the atomic spectra  on the surface of the
star before and after the neutrino ejection pulse provide
information both on the gravitational potential of the neutrino
flux and the neutrino mass, and on the  EP violation.\looseness=-1

Constraints on the EP-violating parameters are also considered in Ref. \cite{NI}
(see second footnote in Subsection B) and in Refs. \cite{Alf,Lam}    (see two last footnotes in Subsection D).
 Possible violation of EP in some exotic cosmological models is mentioned in the end of Subsection G.\\[-5pt]

\noindent\textbf{\vspace*{-0.1cm}G. Manifestations of CPT and Lorentz invariance violation  with `non-standard'
\hspace*{0.6cm} mechanisms in neutrino physics}\\[-10pt]

This Subsection mostly deals with CPT/LI-violating neutrino processes involving
flavor change owing to resonant effects and to `non-standard'\, mechanisms (see Subsection D). The resonant effects
originate in the energy level crossing of neutrinos (and, separately, antineutrinos), at low and high density regions
of supernovas. The `non-standard'\, mechanisms imply, in particular,  quantum decoherence (QD) in the form
of Eq. (10) in the linear formalism mentioned in the middle of Subsection D
 \footnote{These topics are presented in detail in lectures \cite{ma}
 in the framework  of a review on CPT violation in quantum gravity models.}
 and  MDR of Eq. (11) in Einstein's gravity as well as in loop quantum gravity.

Several papers reported an analysis of NO data in two-flavor
approximation based on the possible decoherence effect that
is described by the right-hand side of (10) and parametrized
 by six variables (as mentioned in the second paragraph below Eq. (12)). Earlier
attempts \cite{liu1} of explaining the deficit of solar neutrinos ($\nu_{sol}$),
 as well as atmospheric ($\nu_{atm}$) data, disregarded the requirement of `total positivity' that relates
 these parameters to one another. Then stringent constraints for one of them,  $\gamma = \gamma_0 (E/\rm{GeV})$$^k$
$ (k = -1, 0, 1, 2)$, were found in the simplest single-parameter
case (in the limit when the neutrinos are weakly influenced by
environments); this parameter characterizes the suppression of the
 conventional oscillations term with $\Delta m^2$ through additional
factor exp$(-2\gamma L)$ due to QD. Strong constraints for the parameter $\gamma_0$ were extracted \cite{lisi,Fogl}
\footnote{See also a pessimistic note of Ref. \cite{Adler}.} from the  $\nu_{atm}$
data at 90\% CL for $k = 0,~2,~-1 $, equal to
$3.5\times10^{-23}$ GeV,  $0.9\times10^{-27}$ GeV$^{-1}$, $0.7\times10^{-21}$ GeV,
respectively \,\footnote{See also a note \cite{TOhl} on
the equivalence between two models: averaged over experimental parameters model of NO and  QD
 model, which is used, e.g,  for analysing   the S-K $\nu_{atm}$ data.}.
A detailed analysis of a more realistic case of $k = -1$ \cite{Fogl} on the basis of the result of S-K + К2К
experiments in the channel $\nu_\mu\rightarrow\nu_\tau$ failed to
detect evidence for QD effect; however, it equally not ruled out its presence at $\Delta m^2 = 0$. \\
As for QD as a subdominant effect, still more hard limits on parameter $\gamma_0$ were obtained very recently \cite{FOGL}
from a global fit to solar and KamLAND NO data. They are given at 95\% CL for, respectively, $k = -2, -1, 0,+1, +2$
(here the neutrino energy $E$ is taken in GeV to have all bounds in the same GeV units):
$\gamma_0 <(0.81\times 10^{-28}, 0.78\times 10^{-26}, 0.67\times 10^{-24}, 0.58\times 10^{-22}, 0.47\times 10^{-20})$ GeV.
The last value at $k= +2$ contradicts to the model estimate and is worse than the atmospheric limit.\\
As it was found recently in Ref. \cite{M-S}\,\footnote{See also Ref. \cite{R-N}.}, the
expected sensitivity limits of modern experiments CNGS (CERN-SPS, 400 GeV/c, to OPERA (Gran Sasso), $L\approx$ 730 km)
and T2K (Tokai J-PARC, 40 GeV/c, to S-K, $L\approx$ 730 km), which should be directly compared with above S-K and K2K
results \cite{FOGL}, are as follows, respectively for $k = 0, -1, +2$: CNGS -- $2\times10^{-22}$, $9.7\times10^{-22}$,
$4.3\times10^{-26}$, T2K -- $2.4\times10^{-23}$, $3.1\times10^{-23}$, $1.7\times10^{-23}$.

Earlier fitting of reactor and  short-baseline accelerator
experiments (CHOOZ, CHORUS, E776, CCFR) established upper bounds on
$\gamma_0$ for all values of $k$  at 99\% CL \cite{GSTZ};
constraints in the $\nu_\mu\rightarrow\nu_\tau$ channel were found
to be considerably weaker than those obtained from the
$\nu_{\rm sol}$ data.   In the channel $\nu_\mu\rightarrow\nu_e$
they were, by the order of magnitude, $10^{-22}$ GeV$^2$,
$5\times10^{-22}$ GeV, $5\times10^{-24}$, and $10^{-26}$ GeV$^{-1}$,
 respectively, for $k = -1, 0, 1, 2$ (it appears that the last
two constraints will unlikely be improved using the  $\nu_{\rm sol}$ data); the limits in the channel
$\nu_e\rightarrow\nu_\tau$ are such that the results are more
stringent than those obtained from the $\nu_{\rm sol}$ data only
if $k = 2$: $\gamma_0 \lesssim 10^{-24}$ GeV$^{-1}$.
\footnote{Above-mentioned as well as similar limits are discussed in Ref. \cite{Ancor}
in view of studying QG-induced lepton flavor violation with AMANDA and IceCube detectors
at the South pole.}\\
The same authors carried out a quantitative analysis \cite{GAGO} of
the potential uses of long-baseline accelerator experiments, К2К,
MINOS, OPERA, and of a neutrino factory, in order to discriminate
between ordinary NO and NO due to purely decoherence effects in
$\nu_{atm}$ transitions $\nu_\mu\rightarrow\nu_\tau$.\\
Very recently estimated NO parameters for the MINOS experiment, as it was reported in Ref. \cite{MINOS}, are in agreement
with their current values, provided that observed $\nu_\mu$ deficit and spectrum are due to $\nu_\mu\rightarrow\nu_\tau$
oscillations at $L\thickapprox$ 730 km between two detectors. Best fits to energy spectra in the far detector
disfavor the alternative QD scheme of Ref. \cite{Fogl} at the 5.7 $\sigma$ level.\\
The subsequent work \cite{GA} followed Ref. \cite{GAGO} extended the initial
formalism of Ref. \cite{ben} to the three-flavor system in the
framework of a general approach, which is independent of specifics
of the model of QD interaction between neutrino and
surroundings,  and obtained explicit formulas for NO probabilities
in this case. The authors of this work also studied the correspondence of the
three-flavor analysis to the above-described two-flavor decoherence
analysis  for $\nu_{\rm atm}$. Two qualitative scenarios were
investigated: (1) flavor change in NO due to QD only, and (2)
joint effect of this mechanism and the conventional one.
It was shown that with a simplifying assumption of diagonality of
dissipation matrix in the right-hand side of (10), both versions of
taking QD into account fail to comply with experimental
data if the mixing in the channel $\nu_e\rightarrow\nu_\mu$ or
$\nu_e\rightarrow\nu_\tau$ is included.\\
With a view to improve (by a factor of at least $10^{12}$) the sensitivity for observing
QD by future neutrino telescopes such as IceCube, the authors of Ref. \cite{Hmw}
suggest making use of ultra-high energy $\bar\nu_e$ from cosmic neutron $\beta$ decays. In this case after usual NO
with initial neutrino flavor ratios $1:0:0$ one obtains $0.56:0.24:0.20$ while effects of the QD
in all cases \cite{Ahluw} as well as NO in the case of charged-pion$\rightarrow$muon decays result in $\frac{1}{3}:
\frac{1}{3}:\frac{1}{3}$.

As was shown in Ref. \cite{AGG}, the IceCube detector can strongly
improve the current sensitivity to the $\bar\nu_e$  TeV-energy flux from cosmic-ray
neutron decays in the massive star forming region Cygnus OB2 at $L\approx1.7$ kpc. In a
parametrization  of QD coefficients (similar to schemes of
above-quoted works) in the form $\bar\gamma = \kappa_n(E/{\rm
GeV})^n$, their  upper bounds at 90 (99) \% CL are the
following: $\kappa_{-1}\leq1.0\times10^{-34} ~(2.3\times10^{-31})$
GeV, $\kappa_{0}\leq3.2\times10^{-36} ~(3.1\times10^{-34})$ GeV,
$\kappa_{1}\leq1.6\times10^{-40} ~(7.2\times10^{-39})$ GeV,
$\kappa_{2}\leq2.0\times10^{-44} ~(5.5\times10^{-42})$ GeV,
$\kappa_{3}\leq3.0\times10^{-47} ~(2.9\times10^{-45})$ GeV.
\,\footnote{Previous upper bounds of the QD constants $\gamma_0$ and $\kappa_n$, which are obtained
from simulating atmospheric NO at  $E \gtrsim$ 200 GeV for neutrino telescopes,
 such as ANTARES, were presented in Ref. \cite{Mor}.}

We need to mention in this context that there is an extremely
strong astrophysical constraint on the QD effect:
$\gamma_0 \lesssim 10^{-40}$  GeV at $k = 0$ \cite{Klap}. It is
based on the published estimate of the upper limit of the probability for
recording the fact of NO  from data of the supernova SN1987A at $L\sim50$ Mpc (with taking into
 account all the NO processes): $P(\bar\nu_e\rightarrow\bar\nu_{\mu,\tau})< 0.2$ \cite{Smir}. This constraint
 imposes a very considerable limitation on the expectation of observing the effect in other experiments,
even though the data on NO from active galactic nuclei may amplify it by many orders of magnitude \cite{Klap}.

In the theoretical analysis of LI tests by Planck-scale-MDR of the type (12), a possible sensitivity to
$M_{\rm Pl}$-suppressed effects is expected \cite{Ame} in time delay experiments with very-high energy neutrinos
from gamma ray bursts at observatories such as ANTARES. This case, with and without the emergence of a preferred
inertial frame, is not affected by the delicate problem, discussed by the author, of a deformed law of energy-momentum
conservation.\\
The author of paper \cite{Chr} argues that for cosmogenic neutrinos at very high energies of $10^{18} - 10^{21}$ eV
the Planck-scale corrections to NO length, that are suppressed even by $M^7_{\rm Pl}$ factor in QG-MDR of Eq.
(12)-type, may be detected with telescopes such as IceCube and ANITA.

In the paper \cite{Hoo} a review is presented on the theoretical foundation and motivation for studying CPT/LI
violation with high-energy neutrinos originated from cosmically distant decays of the charged pions and the neutrons.
The goal is to obtain signatures for identifying the violation effect by confronting observed neutrino flavor ratios
$R_q ~(q = \nu_e, \nu_{\mu}, \nu_{\tau}$) to predicted ones, which were calculated and displayed for two simplified
illustrative scenarios of the energy behavior of $R_q(E)$ at 10 GeV $\lesssim E \lesssim 10^8$ GeV
with MDR of the Eq.(12)-type when substituted $M_{\rm Pl}$ instead of $M$.
 When considering Lorentz-violating case based on a general off-diagonal hamiltonian formalism with one real parameter
explicitly proportional to $E^2$ or $E^3$,  the initial set of $R_q $ for pionic origin
is radically changed above the energy threshold of the effect at $E_{thr} \thicksim$ 1 TeV or $\thicksim10$ PeV
(1 PeV$=10^{15}$ eV), respectively, while for neutronic origin it alters there to a smaller degree.
In the case of CPT violation, a second approximation scheme is considered, with one nonzero QD parameter only,
in which, for simplicity, only one $L$-dependent term $e^{L\delta}$ was left, after averaging to zero the sin and cos terms
 over large distances. Then, in a model with $\delta = (E/10\,000 ~\rm{GeV})^2$/$L$
  all the $R_q(E)$ came to the same value, $\frac{1}{3}$, -- gradually for pionic origin while dramatically for
  neutronic one. In conclusion, the signatures are discussed  of QD effects on neutrino events observed in high-energy
telescope, such as IceCube and KM3.

Because the first physical data from neutrino telescopes can be obtained even in the $\nu_{atm}$ exposition,
authors of Ref. \cite{Morg} proceeded to study probing LI but in the (12)-type scheme of MDR as in Ref. \cite{Hoo}
at $u \geq 2$ (in particular,  of DSR-type) and simulated the ANTARES sensitivity, as in Ref. \cite{Mor}. In further
discussion they used  upper bounds on Lorentz-violating parameter estimated, when additional term in
NO phase is proportional to $E^{u'}$, in two neutrino mass cases: at $\Delta m^2 \neq 0~ -~2.9\times10^{-24}, ~u' = 1$;
$~2.9\times10^{-35}$ eV$^{-1}$, $u' = 2$; $~6.9\times10^{-46}$ eV$^{-2}$, $u' = 3$; and at $\Delta m^2 = 0~
 -~8.2\times10^{-25}, ~u' = 1$; $~1.0\times10^{-35}$ eV$^{-1}$, $u' = 2$.

 Meanwhile, recent consideration of MDR of the (12)-type and its influence on $R_q $
of high-energy neutrinos originated from distant cosmic sources shows \cite{Bust} that even in this case the sensitivity
 of future IceCube and ANTARES telescopes will not ensure the identification of the CPT/LI-violating effects through
 variation in the averaged $R_q$ observed .\\
Possible tests of LI violation when they sensitive  to Planck scale physics
due to Eq. (12) depend on neutrino energy extension: in atmospheric $\nu$ experiments the PeV energies
 will be  accessible  while in gamma ray bursts (GRB) the values about exo-eV (1 EeV $=10^{18}$ eV)
are expected. From this point of view, in Ref. \cite{GGH} the problem of observing GRB neutrino events
of three flavors with its time delays and energies is reconsidered in detail for the case of IceCube-type detectors.

Another source of CPT violation is the interaction of fermion spin
with the spin connection of the external gravitational field in
Einstein's theory provided its sign is not reversed under  CPT. The
contributions of this interaction to the energy for  Dirac
neutrino and antineutrino are of opposite signs. So, MDR are appeared
with helicity energy gap for $\nu$ and $\bar\nu$. This fact results
in unequal number densities of $\nu$ and $\bar\nu$: in the early
Universe -- due to their scattering on primordial black holes \cite{Sin}
and in axially symmetric cosmological solutions  \cite{Sin, Mukho},
 in today's epoch -- via scattering on rotating black holes
\cite{Sin,Muk}. In the general case of CPT/LI violation with Dirac and Majorana
masses \cite{bar3}  and taking into consideration recent  works on NO in gravitation field,
authors of Refs. \cite{Muk6,Muk7} and \cite{Muk9} argue, that strong axial spin-gravity
 coupling results also in  neutrino--antineutrino oscillations without flavor changes,
 which have no dependence on $m$ and $E$ \cite{Muk6,Muk7,Muk9}.  The $\nu-\bar\nu$ asymmetry \cite{Mu}
and $\nu\leftrightarrow\bar\nu$ oscillations in the anisotropic
phase of early Universe may lead to lepto- and baryogenesis \cite{Muk7,DMD} (and, possibly, contribute to r-process
nucleosynthesis in supernovae \cite{Muk7});
 the $\nu-\bar\nu$ mixing influences the $2\beta0\nu$-decay rate \cite{Muk9} as the mass of the flavor state is modified.

Some papers, in which authors work within loop quantum gravity \cite{AMU}-\cite {LaSi}, treat
MDR of Eq. (11) too. Besides, the theory assumes
the existence of an intermediate scale $\mathcal{L}\gg L_{\rm Pl}$
that separates the lengths $d\ll\mathcal{L}$ at which the loop
structure of space manifests itself, and the lengths $d\gtrsim
\mathcal{L}$ at which flat classical geometry is regained.
Detailed investigation showed \cite {Alf, AMU} that in the
framework of this approach the function $F$ in (11) for the
neutrino in vacuum is in general parametrized by nine constants of
different degrees of suppression (compared to unity) through a
factor $(L_{\rm Pl}/\mathcal{L})^{3\Upsilon + 2}$
 ($\Upsilon\geq 0$ is an additional phenomenological parameter that is
possibly a function of energy), with the scale  $\mathcal{L}$
defined for two scenarios ($\mathcal{L}\sim 1/E$ and
$\mathcal{L}\sim$ const); the linear in $p$ additive term in $F$
includes the sign `$\pm$' in accordance to the helicity. The
authors of Ref. \cite {AMU} analyzed in detail the possibility of
extracting information (with evaluating the restrictions on the
parameter $\Upsilon$ based on the  $\nu_{\rm atm}$ data) on two
characteristics of observable (in principle) effects of cosmic  GRB
as they are accompanied with powerful ejection of
massive neutrinos with $E \sim 10^5 - 10^{10}~$ GeV: (1) by the signal
delay time for various neutrinos  in comparison with the
light signal, found to be of the order of $(EL_{\rm Pl})L/c \approx 10^4$ s;
 (2) by the $E$-dependence, $L_{\rm osc}^{-1}\propto E^2 L_{\rm Pl}$, which differs from that
discussed in Ref. \cite {fog} (see the beginning  of Subsection F).\\
On the basis of the above formalism with working set of
parametrization constants and by comparing theoretical results with
NO data and with the spectrum of cosmic rays of extragalactic
origin, the following problems were also considered: the energy
dependence of NO length \cite {Lamb}, constraints on the
intermediate scale ($\mathcal{L}\gtrsim 10^{-18}~$eV$^{-1}$ \cite
{Lamb,lamb}) and the working constant \cite {lamb}, and a novel
mechanism \cite {LaSi} for generating the primary cosmological
asymmetry of the Universe originating from the  density difference
of neutrinos and antineutrinos caused by the above distinction in
signs for the linear-in-momentum contribution to function $F$ in Eq. (11).

According to some particular QG models (see, e.g., Refs. \cite {John,AMU}), the lower order
in $E_{QG}$ energy-dependent corrections of the general type to
Lorentz-violating MDR of Eq. (11) may originate from the dimension $d
= 5$ operator while $d = 4, ~6$ operators are $E$-independent. The
paper \cite {chou} is the first study to date in which the time delay, $\Delta t$, for massive neutrinos from
GRB is calculated with correctly accounting the time dependence of the Hubble constant due to matter effects as well as
 dark energy effects. The analysis for $F \propto \pm E^2(E/E_{QG})$
 was carried out for observable values in the planes $\Delta t$ {\it vs} $E$ in the wide energy interval,
$\Delta t$ {\it vs} $m/E$ and $E/E_{QG}$ for redshift $z = 0.01 - 10$ and for three different cosmological models.\\
Very recently, after considering the stochastic space-time foam models, more generic than in Ref. \cite{MaSar},
with small random fluctuations about flat Minkowski background, the authors came to conclusion \cite{AlMav}, that the current NO
data do not have the sensitivity to test such the models; only high-energy neutrinos from GRB can constitute
sensitive probes of QD \cite{Hmw,AGG}.

 For the deformation parameter in a non-commutative fields scheme with massless-type MDR  (mentioned at the very end of
 Subsection D), the following bounds are obtained from LMA scenario at mixing angles
 $\theta \sim \pi/4$ on the basis of atmospheric and solar neutrino results, correspondingly:
 $|\alpha_{23}| < 10^{-22}, |\alpha_{12}| < 10^{-17}$ \cite{Ari}.

For exploring quantum gravity-induced violation of LI in Ref.
\cite{JP}, the MDR (11) with $F \simeq\pm E^2(E/\xi M_{\rm Pl})^u$ $(\xi$
is scale parameter, $u = 1, 2, 3,...$) were used. The authors consider, for $u = 1$ and 2,
the feasibility of detecting the  effect, with the presence of background muon events from $\nu_{atm}$,
on the basis of estimating $\Delta t$ between neutrino and low-energy photon signals from cosmic GRB
while $\Delta t$ of order of hours are expected for 100-TeV $\nu$'s at redshift $z=1$ and $u = 1$.\\
MDR of the similar type are considered in Ref. \cite{Di G} to put
bounds on LI-violating neutrino coefficient $\tilde\alpha_{e(\mu)}$
and mass scale $\mu_e$ by using relations $E^2 = p^2 + 2\mu_e|p| ~~$
and $E^2 = p^2 + \tilde\alpha_{e(\mu)}|p|^3/M_{\rm Pl}$ (as the $u =
1$ and $u = 3$  cases of MDR (12) for $M_{\rm QG} \rightarrow M_{\rm
Pl}$ within the framework of the doubly special relativity (DSR)\;
\footnote{Besides a discussion  around Eq. (12) on non-linear type
MDR, see also Ref. \cite{matt} at the end of Section 2 where the role
of the DSR is presented in modern tests of LI.}) for its
substitution to formulae for the widths of the $\pi_{e(\mu)2}$ decay
and the neutron $\beta$ decay. The authors give also examples of MDR in other interesting models.\\
 When studying a generalized approach for deducing the equation of motion and
the dispersion relation of a propagating fermionic particle, the author of Ref. \cite{Bern} recovered the corresponding
 VSR formulas of Ref. \cite{CoG}. Then, the new LI-violating VSR-dynamics was applied to the tritium $\beta$-decay
in order to compare the modified Curie-plot with the end-point spectrum data as was pointed out
in Ref. \cite{CoG}. \\
In subsequent work \cite{BB}, with usual approach for obtaining LI-violating neutrino dynamics similar to VSR,  new
MDR are found via a specific transformation combined boosts and rotations with the aid of a light-like preferred vector,
while the Lorentz algebra holds. Then an effective neutrino mass effect in this preferential direction scenario  also
confronts the MDR parameters with data for tritium end-point spectrum.
\footnote{Additional corrections to this spectrum via de Sitter symmetry breaking in VSR are
discussed in Ref. \cite{Alvar}.} Analogous suggestion
was made early in paper \cite{BMP}, with using MDR of DSR-type \cite{MaSm} which appears in a study of LI for mixed
neutrinos (see mentioning these works in discussion on non-linear MDR in Subsection D below Eq. (12)).\looseness=-2

As it has been suggested about ten years ago, space-time foamy structure in quantum gravity may imply  LI breaking
with  MDR (see discussion around Eq. (11))
that leads to  deviations of velocities (sub- or super-luminal) for massless  neutral particles from speed of light.
In this respect, the Ref. \cite{Eetal} gives the relevant limits on linear or quadratic Lorentz violation when
observing energetic neutrinos from SN1987A in the form $\delta v = (E/M_{QGl})^l$, with $l = 1, 2$.
Based on statistically poor data sets from KamiokaII, IMB and Baksan detectors with event lists of neutrino energy
and arrival time and using minimal dispersion method, the authors obtained the following results at 95\% CL for
sub-(super-)luminal cases, respectively:
$M_{QG1} > 2.7(2.5)\times10^{10}$ GeV, $~M_{QG2} > 4.6(4.1)\times10^{4}$ GeV.  For the possible  registration
of future galactic supernova at 10 kpc in the S-K, Monte Carlo simulation gives rise:
$M_{QG1} > 2.2(4.2)\times10^{11}$ GeV, $~M_{QG2} > 2.3(3.9)\times10^{5}$ GeV.
Potential sensitivity (in the case of future improvements) of the OPERA detector
(long-baseline beam from CERN to Gran Sasso) as estimated in Ref. \cite{Eetal} exceeds the
$M_{QG2}$ limit: $M_{QG2} \sim 7\times10^{5}$ GeV.

With the goal of using supernova neutrinos for testing CPT symmetry, authors of Ref. \cite{MU} consider
feasible level-crossing patterns of the resonant flavor conversion in SN progenitors and estimate fluxes
of three neutrino species -- $\nu_e, \bar\nu_e,$ and $\nu_x$ as collective notation for $\nu_{\mu}, \bar\nu_{\mu},
 \nu_{\tau},$ and $ \bar\nu_{\tau}$. These authors concentrate on difference between $m_{\nu}$ and $m_{\bar\nu}$
 patterns (as well as on $\theta_{13}$ and $\bar\theta_{13}$ ones) and analyse how flux features of six working
 patterns can be discriminated during the observation of future galactic supernovas.

Finally, a separate problem discussed in the literature stems from
attempts at describing the closeness  of the observed neutrino
masses and the energy scale of the dark energy, which determines the
acceleration of the cosmological expansion of the Universe. The
authors of Refs. \cite{GBMaV,mavro}, extending and specifying the general
line of their above-mentioned  work \cite{GBMav} (besides a new
interpretation of all NO data including LSND \footnote{Issues on LSND anomaly see in Section 5
 (Subsections I, J and K).})  continue also the study  of possible
model approaches to interpret the cosmological $\nu-\bar\nu $
asymmetry and to obtain a meaningful evaluation of the vacuum
energy (the so-called dark energy or the cosmological term)
\footnote{The authors proceed from the non-equivalence of the flavor-state
and of mass-state neutrino vacua; this results in non-trivial
contribution to the cosmological term owing to the mixing effect
itself \cite{BLAS}.}  caused by mixing of neutrinos through
QD. The starting point is the assumption  \cite{GBMaV,mavro} that
QD effects (caused by the interaction of the neutrino with
the foam structure of the vacuum (see Subsection D)) contribute to
the terms of hamiltonian in Eq. (10) for the evolution of the density matrix and result in the emergence of
effective mass shifts, by analogy to the MSW effect in a medium.
\footnote{In Ref. \cite{GBMaV} a possible scenario is suggested in which, provided that only in antineutrino sector
there is CPT-violating QD, due to charge conservation microscopic charge black hole--anti-black hole pairs in
space-time foam can produce mostly $e^+e^-$ (but rather not $\mu^+\mu^-$) pairs. Then, if, e.g., positrons are absorbed by
event horizons during evaporation of anti-black holes, stochastically fluctuating electron density excess is originated.}\\
Note that the application of the idea of a generic origin of the
neutrino mass and the dark energy in one form or another is typical
for a lot of works  studying the possible interactions between
neutrino (taking account also of varying mass) and  dark energy.
Different models with spontaneous and dynamical baryo- and
lepto-genesis induce CPT and LI (and in some cases also EP) violation \cite{MLi,PGu,far}
  \footnote{For reviews of another analogous schemes see Ref. \cite{NI} (which discusses the cosmological
 birefringence in the end of second footnote in Subsection B), Refs. \cite{PGu}, and talk \cite{XING};
 in view of  Refs. \cite{far} see also references therein and  the discussion in paper \cite{Pecc}.},
that can be tested in future NO experiments. Among earlier works in this field,
we mention Ref. \cite{HORV} where EP violation in a scheme  of the cosmological `quintessence' is considered.\looseness=-2

Completing the discussion of the general consequences which relate
 CPT with neutrino physics, we need to remind the
reader that the CPT non-invariance of the theory results first of
all in the independence of the  СР- and Т-violation effects.\\[-5pt]

\noindent\textbf{\large 5. Interpretations of neutrino oscillations based on
CPT violation}\\[-10pt]

In what follows we consider those various (different in principle)
manifestations of CPT-odd effects that may originate NO in terrestrial experiments. \\
We begin (see Subsection H) with publications in which no true
(fundamental) violation of CPT invariance is assumed in the theory
but in which oscillations are treated in the conventional (for the most part,
CPT-non-symmetric) medium.\\
 In Subsection I  attempts are discussed to
interpret NO not by deriving them from the fact that neutrinos are
massive, but on the basis of LI and CPT  violation in the theory
(similarly to those we discussed in Section 3, including those
originated from  EP violation in gravity theories).\\
 Some hypothetical ways of explaining the LSND anomaly are given in
Subsection J. Subsection K discusses papers aimed at obtaining
constraints on the parameters of the fundamental CPT violation in
the case when neutrino and antineutrino parameters are distinct (or
at least $m_{\bar \nu} \neq m_{\nu}$). We note  here once more, that the
authors of the papers with $m_{\bar \nu} \neq m_{\nu}$ assume groundlessly  that LI is conserved
(see  discussion of this problem in the beginning of Section 3). \\[-5pt]

\noindent\textbf{H. False CPT-odd effects in  matter}\\[-10pt]

A number of papers \cite{be1}-\cite{xin} analyzed  false CPT-odd
effects caused by  medium influence on NO while conserving the
CPT invariance of the theory; these effects result, first of all,
in nonzero asymmetry of the probabilities of diagonal transitions,
 $\Delta P^{\rm CPT}_{\alpha \alpha}\equiv
P_{\nu_\alpha \rightarrow \nu_\alpha} - P_{\bar \nu_\alpha
\rightarrow \bar \nu_\alpha} $, i.e., the difference of survival
probabilities of  neutrino and antineutrino of a given
flavor, and in similar asymmetries, $\Delta P^{\rm CPT}_{\alpha
\beta}\equiv P_{\nu_\alpha \rightarrow \nu_\beta} - P_{\bar
\nu_\beta \rightarrow \bar \nu_\alpha} $, for non-diagonal
transitions. For instance, in very-long-baseline experiments for
terrestrial and atmospheric neutrinos, the dependence was
demonstrated \cite {be2} of $\Delta P^{\rm CPT}_{\mu \mu}$  on
energy and angle $\theta_{13}$, i.e.,  on a small matrix element
$U_{e3}$ which characterizes the connection between
atmospheric and solar ranges of $\Delta m^2$ in NO.  As it was shown there is the
familiar resonance effect in atmospheric neutrinos that manifests
itself clearly  at $L\gtrsim$ 7000 km through the interactions of
$\nu$ and $\bar{\nu}$ in the earth's mantle and crust; in this
case the measurement of the CPT-odd asymmetry will provide an
information on $\theta_{13}$ and on the sign of the corresponding
difference of squared masses. In Ref. \cite{xin} a calculation of
$\Delta P^{\rm CPT}_{\alpha\alpha}$ for reactor-based oscillation
experiments on long baselines from 730 to 3200 km was also given.\looseness=-2

A detailed set of approximate analytical formulae for $\Delta
P^{\rm CPT}_{\alpha \alpha}$ and $\Delta P^{\rm CPT}_{\alpha \beta}$
in a medium with an arbitrary density distribution was presented in
works \cite{jac}. Particular cases of constant density and stepwise
density distribution are also considered, the latter corresponding
to NO in  accelerator and reactor experiments at long baseline as
well as at future neutrino factories.\\
 Estimates were obtained (numerically and on the basis of perturbative and low-energy
approximations) for about a dozen of  experiments -- current and in
preparation -- for the indicated CPT-odd differences. Also shown
graphically are the energy and baseline  dependencies of  the effect
for three more efficient accelerator experiments, KamLAND, BNL NWG
and NuMI, for which numerical values of   $\Delta P^{\rm CPT}_{ee}$
and $\Delta P^{\rm CPT}_{\mu e}$ are $-0.033$, 0.032 and 0.026,
respectively \cite{jac}.\looseness=-2

In Ref. \cite{Ble} a general phenomenological approach to NO (in
matter, in particular) is discussed when different damping effects
are introduced to $\Delta P^{\rm CPT}_{\alpha \beta}$. A three
flavor analysis is presented to distinguish among damping signatures
the decoherence-like effects in short- and long-baseline
reactor and neutrino factory experiments.\footnote{While in Ref. \cite{Ble} NSI effects in NO are considered
on probability levels and based on `damping' signatures, future neutrino factory experiments in the subsequent
work of the same authors \cite{Blen} are parametrized and analysed on hamiltonian level. One of the aims of the work
is to find how the precision measurement of NO parameters may be adjusted only due to NSI effects when this effects
can be distinguished from experimental inconsistences such as, e.g., the neutrino decoherence.}
 The information given may help to preserve from faking effects when one interprets, e.g., QD
as a wave packet decoherence \cite{FS} or as NO to sterile neutrinos.

False CPT-odd effects are discussed also in Subsection J when explaining the LSND anomaly \cite{GBMaV,mavro}
 in connection with disentangling genuine CPT/LI violation due to QD from "fake" \, effects
in matter \cite{TOhl,jac}.\\[-5pt]

\noindent\textbf{I. Neutrino oscillations due to violation  of Lorentz invariance}\\[-10pt]

The last part of the talk \cite{K}  cited in Subsection B contains,
besides a discussion of application of the Extended Standard Model (SME)
to neutrino physics, a description of qualitative features of the
simplified two-parameter model (`bicycle' model \cite{KM}) and an analysis of its compatibility
with the data on atmospheric and solar neutrinos. It is noted that
it is currently difficult and would most likely be wrong to exclude
the possibility of describing the observed oscillations due to LI
and CPT violation  instead of assigning mass to the neutrino.
Therefore, to explain   NO exclusively in the framework of CPT-
and LI-violating models, detailed analyses of data of the concrete experiments are  required.

Three articles of Kosteleck\'{y} and Mewes \cite{KM,KoM}, in  two of which the general formalism presented \cite{KM}
for analysing violation of CPT and/or LI in the framework of SME, consider also possibilities of this scheme
to interpret the LSND anomaly by the effects of unusual energy and directional dependence and $\bar\nu-\nu$
mixing. The third paper \cite{KoM} contains the results of studying LI violations
in short-baseline NO  in the framework of the formalism of the SME (see formulas (6) and (7)).
Using general form of parametrization of these effects \cite{K, KM} allowed
to combine  without contradiction the descriptions of the  accelerator experiment LSND
 \footnote{In addition, attempts to explain the LSND anomaly are discussed
  in Subsections J and K.} and reactor
 experiments CHOOZ and KARMEN, due to both a nonstandard functional behavior of relevant terms
 on $E$ and a dependence on the neutrino beam direction.\\
  The two-flavor analysis of LSND data that
 covered a large number of parameters (the terms  $a^\mu L$ и $c^{\mu\nu}LE$  in (6) corresponds to taking into
 account 41 degrees of freedom, including the sidereal-time dependence on direction) yielded a nonzero
quantity    for a combination of coefficients that give the value of
the LI violation. It is found to be  $(3\pm1)\times 10^{-19}~$ GeV, which is characteristic of effects
at the Planck energy scale  and is based on the measured probability over a sideral day
$\langle P(\bar \nu_\mu \to \bar \nu_e)\rangle \simeq (0.26\pm0.08)$\%\cite{KoM}.

Thereupon, the authors of Ref. \cite{BMW2} tried to use a five-parameter scheme of CPT/LI violation as a linear
combination of generalized four-parameter bicycle model with the dependence of coefficients
$(a_L)_{e\mu} = (a_L)_{e\tau}$ on preferred direction of $a_L$ interaction and of original two-parameter
bicycle model of Ref. \cite{KM}  without such dependence. It was shown, however,  that this scheme with $m=0$
contradicts atmospheric, long-baseline and reactor neutrino data in total.

Meanwhile, very recently a searches for sidereal effects in  accelerator $\nu_{\mu}$ data with the MINOS
 Near Detector were failed \cite{adam} as well as attempts to discover analogous periodic variations with respect
to the Sun-centered frame in any previous terrestrial NO experiments.\\
At expected suppression
factor of $M_W/M_{\rm {Pl}} \sim 10^{-17}$ upper bounds on the CPT/LI violation coefficients in the SME
analysis are found to be at the level of 0.01--1 \% \cite{adam} while  the LSND data are consistent with no sidereal dependence
 \cite{auer}.

Nevertheless, the further elaboration of SME-based models pointed out \cite{KKT} that all the NO experiments,
 including solar, atmospheric and reactor data, are described
  by simplified effective  CPT/LI-violating hamiltonian  with three degrees of freedom,
provided the high-energy pseudomass coefficient comes from one Lorentz-violating see-saw mechanism while
the second see-saw in this `tandem' model works at low energies to incorporate $L/E$ dependence of the KamLAND
and features of the LSND. The global NO data, adding also the LSND result, consistently answer  the model at
   $\breve{a} \equiv(a_L)_{\alpha\alpha'}= -2.4\times10^{-19}$ GeV  (here $\alpha \equiv\alpha'= e, \mu, \tau$),
$\breve{c} \equiv -\frac{4}{3}(c_L)_{ee}= 3.4\times10^{-17}$, $\breve{m}^2/2  \equiv (m^2)_{\tau\tau}/2
= 5.2\times10^{-3}$ eV$^2$
in the direction-independent approximation and without MSW matter effects within the Sun.\,
\footnote{A brief information is given in Ref. \cite{MatM} on LI-violating models (bicycle and tandem)
which may promote  the joint explanation of all the NO data (including LSND).}\\
Another three-flavor direction-independent approach but without CPT violation is explored in Ref. \cite{dGG},
where LI-odd effects are very small for experiments with solar and KamLAND neutrinos while atmospheric neutrinos
are described, in fact, as in the two-flavor scheme. In order to accommodate the LSND data, only one relevant  neutrino
was coupled to LI violation provided that the non-trivial energy fine-tuning MDR are taken on. \\[-5pt]

 \noindent\textbf{J. CPT non-invariant `ether', quantum decoherence, and  LSND anomaly}\\[-10pt]

Another idea tested for the interpretation of neutrino experiments
was the CPT-non-invariant `ether' \cite{bar2} acting as a dense medium and
 creating interaction potential of opposite signs (and, correspondingly, different {\it effective} masses) for
neutrino and antineutrino. As it was shown in Ref. \cite{gou}, a two-flavor analysis for this model
with introducing Lorentz-non-invariant {\it effective} operators cannot
solve  problems of the solar neutrino deficit and of the anomalous result of LSND.\\
However, in principle, the increased number of fitting parameters
and their non-standard dependence on energy provide a feasibility of
describing NO data, including LSND.

 In  Subsection I we quoted the possibility of interpreting experiments on a short
baseline only \cite{KoM} via LI-violating effect in the Extended
Standard Model. Such approach,  assuming also \cite{GBMav} non-identical QD parameters
for  neutrino and antineutrino due to  the strong CPT violation
but with $m_\nu = m_{\bar \nu}$, made it possible  to successfully
fit all the available NO results. The application of the three-flavor
analysis, of simplifying assumptions on parametrization of the QD effect (which works in this particular model
only in the antineutrino sector and is described by two quantities that are
directly and inversely \footnote{The terms proportional to $E^{-1}$ were attributed
to usual matter effects.} proportional to $E$), and of the conventional
NO mechanism with $\Delta m^2$ allows to explain  \cite{GBMav,GBMaV,mavro}
\footnote{To survey problems under discussion see also Section 13.2 of the White paper \cite{White} and review \cite{StVi}.}
the LSND anomaly. Important points, which contribute to this explanation,
are the suggestion \cite{GBMaV,mavro} that the observed neutrino mass differences are originated from a
stochastic space-time foam and  the appearing an exponential-suppression factor multiplying the oscillation term
provided that one takes into account the difference in $L$ for different experiments.\\
In contrast to  above purely phenomenological model \cite{GBMav,mavro}
of quantum gravity-induced decoherence, in Ref. \cite{BMSWL} authors
presented also a complete, mathematically consistent, QD scheme of
three-flavor mixing to describe all NO data (including LSND plus KamLAND spectral distortion results)
assuming the same parameters for $\nu$ and $\bar\nu$. One of significant conclusions
from successfully performed fitting, as it seems, is fairly unexpected interpretation
of it: at obtained values of parameters it is more naturally to connect NO not with QD
but rather with different possible origin -- usual uncertainties in the (anti)neutrino beam energy
that appear as "fake"\, effects \cite{TOhl}. Therewith, the fitting leads to the large QD effects with so non-naturally
suppressed factors that may only be tested at higher energies \cite{BMSWL}.\\
Note in addition, that in the review talks \cite{mavrS,MavrS} (which followed above works) with attempts to explain
all the NO data due to CPT-violating effects, the authors now exclude foam-induced QD as the sole cause of the fitting
result obtained \cite{mavrS} and hope rather upon future experimental `solution' of the LSND anomaly \cite{MavrS}.\\
Nowadays  summary of reconciling all the NO results on the base of dumping QD-type effects is presented in
Ref. \cite{FSchS}, the main goal of which is to suggest the consistent three-flavor scenario with the same suppression in
neutrino and antineutrino sectors. The model assumes the fast energy dependence of the QD parameter $\gamma$ which
works only in the $\nu_3$-mass state (and explains the LSND $\bar\nu_e \rightarrow \bar\nu_{\mu}$ events) but
does not affect $\nu_1 -  \nu_2$ mixing while it  can strongly influence supernova signals.

The LSND experiment and its result themselves deserve some additional remarks. We know (see, e.g.,
 reviews \cite{ak,McK}) that the data on the deficit of solar and
atmospheric neutrinos (confirmed in a number of experiments) were
successfully explained in the scheme of three-flavor mixing; this
approach operates only with  two independent differences of squared
masses  $\Delta m^2_{ij}$ ($i,j = 1, 2, 3$). Therefore the
indication in favor of the third value of $\Delta m^2$ that was
obtained in the LSND experiment requires a modification of the
scheme through incorporation of sterile neutrinos (i.e., by adding
$i,j > 3$) or through a radical increase in the number of its free
parameters in the case of CPT breaking (see the beginning  of Subsection B).\\
The LSND experiment \cite{LSND}  searched for $\bar\nu_e$ events
originating in decays of positive muons. These muons  were produced in decays
of stopped pions generated in interactions of protons from the LAMPF
linac. An analysis of the data led  to a conclusion \cite{LSND} that
a non-diagonal transition  $\bar \nu_\mu \rightarrow \bar \nu_e $ at
such third value $\Delta m^2_{\rm LSND} \sim $ 1 eV$^2$ was present.
This result was only obtained in a single experiment and was never confirmed by similar measurements at the KARMEN2
experiment \cite{KAR} (see also the negative conclusions in Ref. \cite{ch}
 on a joint analysis of these experiments and  of searching for oscillations $\bar \nu_\mu \rightarrow \bar \nu_e $
in the accelerator experiment NuTeV \cite{NuT}).\\
The current  accelerator experiment MiniBooNE (FNAL) \cite{mnb}  aims to test the LSND result.
 The refutation of this result would mean that there is no need to introduce sterile neutrinos
 \footnote{In the past, in Refs. \cite{strum}  (see, also, review \cite{MSTV})
exotic scenarios were used  to interpret LSND data in conjunction
with all other NO results: with the fourth sterile neutrino and
nonobligatory equality of $\nu$ and $\bar\nu$   parameters (with
$m_\nu \neq m_{\bar{\nu}}$, in particular). It was shown in the
CPT-violating case that while mass spectra of active + sterile
neutrinos of the types (3 + 1) and (2 + 2) were possible for
$\bar\nu$, only the (3 + 1) scheme is valid for $\nu$; in
CPT-conserving scenarios the (3 + 2) models fit data better than (3
+ 1). Then, after the first results from the MiniBooNE experiment
\cite{Mini}, which found no evidence for the LSND $ \nu_\mu
\rightarrow  \nu_e $ signal, the three-flavor global NO fitting
\cite{MaSch,Schw} disfavored models with one and more than one
sterile neutrino without CP-violation effects. The present status of active + sterile mixing models
(up to 5-$\nu$ and 6-$\nu$ schemes) is also summarized in Ref. \cite{GGMM}.}
or a hypothetical inequality of the neutrino and antineutrino masses.  As for MiniBooNE {\it vs} LSND,
 the very last status (prior to 2009) was presented in Ref. \cite{mini}
in which a joint data analysis of these experiments with Bugey and KARMEN2 obtained a maximal compatibility
of 3.94\% within two-neutrino approach.\footnote{As multiple cycles of NO have not yet been found and assuming,
as usual, only two mixing channels, the definitive conclusions on the compatibility among LSND, KARMEN2 and MiniBooNE
data may essentially depend \cite{gold} on taking into account multichannel mixing, e.g., intermediate sterile-neutrino
channels.}
 However, any confirmation of the LSND anomaly may attract additional attention to
using the simplest, even though theoretically unfounded, model for
interpreting oscillations via $m_\nu \neq m_{\bar \nu}$  (see Subsection K).\\[-5pt]

\noindent\textbf{\vspace*{-0.1cm}K. Antineutrino- \textsl{vs} neutrino-oscillation parameters;\\
\hspace*{0.6cm}LSND-anomaly models with \textsl{m$_{\bar \nu}\neq ~$m$_{\nu}$}}\\[-10pt]

As for   the schemes with $m_\nu \neq m_{\bar \nu}$ (i.e., $m_{\alpha}\neq \bar m_{\alpha}$, where  the mass-state
index $\alpha = 1, 2, 3$), it was shown, for example, that in neutrino factory experiments
sensitivity at the level of $|\bar{m}_3 - m_3|\lesssim
1.9\times10^{-4}$  eV can be achieved \cite{bil2}.

In Ref. \cite{str} data fitting was carried out  for the S-K + К2К
experiments at $m_\nu \neq m_{\bar \nu}$ in the range of
atmospheric squared mass difference.   The first parameter in the
$ \Delta m^2_\nu~ vs~ \Delta m^2_{\bar \nu}$ diagram proved to be
constrained as in the case of CPT preserved while the
second one, in contrast, showed the allowed values larger by about an
order of magnitude.

NO experimental data were also analyzed in order to evaluate $
\delta = \Delta m^2_\nu - \Delta m^2_{\bar \nu}$ \cite{mu}. The
author cites the result $(- 7.5\times 10^{-3}~\mbox{eV}^2 < \delta <
5.5\times10^{-3}~\mbox{eV}^2$) presented by the S-K collaboration
at the conference ICHEP-2002. This result is based on studying the flux of
atmospheric neutrinos and points to its dependence on the
assumption that mixing is maximal and identical for $\nu$ and
$\bar\nu$. However, a more detailed analysis of the most
 recent SNO data using the  MSW  mechanism  inside
the solar matter and  the information on the deficit of reactor
antineutrinos in the KamLAND experiment \cite{kam} yields a better
constraint on $ \delta $ for any pair of flavors and for the case
when $\Delta m^2_{\nu,\bar\nu}$ may have different signs:
$|\Delta m^2_\nu - \Delta m^2_{\bar \nu}|<
1.3\times10^{-3}~\mbox{eV}^2$ (90\% CL) \cite{mu}.

As for the  data on possible CPT non-conservation in the neutrino and antineutrino sectors, they  are  given
in the review analysis \cite{BGP}, where this situation  is
represented graphically by comparing the allowed areas of NO parameters before and after the Neutrino-2004
conference (including the most recent KamLAND results presented to the conference \cite{Araki}).
After the Neutrino-2004 the area for reactor $\bar\nu_e ~$ was shrunk
like one for solar $\nu_e$,  confirming the CPT symmetry in  $\nu_{e}$ sector.\\
Very recently in Ref. \cite{AFM} the authors investigated the sensitivities of future NO experiments
for measuring $\Delta m^2_{ij}$ and $\theta_{ij}$ independently for neutrinos and antineutrinos.
Expected sensitivities of neutrino factories to the atmospheric, (anti)neutrino parameters
were also updated. Present and future bounds for solar, $\nu$ and $\bar\nu$ parameters,
which expected  at $3\sigma$ level in a dedicated $\beta$-Beam facility, in combination with a SPMIN
reactor experiment, are as follows (see  below  reduced variant of the complete Table 1 of Ref. \cite{AFM}).

\begin{center}
\vspace*{-0.3cm}
{\renewcommand{\arraystretch}{0}
\begin{tabular}{|c|c|c|}
\hline
       \strut quantity &{present bound}& {future ($\beta$B 750 km)}\\\hline
\hline \rule{0pt}{2pt}&\\
       \strut $\mathrm{|sin^2\theta_{12} - sin^2\bar\theta_{12}|}$ & 0.3& 0.14  \\
\hline \rule{0pt}{2pt}&\\
       \strut $\mathrm{|sin^2\theta_{13} - sin^2\bar\theta_{13}|}$ & 0.3& 5.7$\times 10^{-4}$ \\
\hline \rule{0pt}{2pt}&\\
       \strut $\mathrm{|sin^2\theta_{23} - sin^2\bar\theta_{23}|}$ & 0.45 & 0.044  \\
\hline \rule{0pt}{2pt}&\\
       \strut$|\Delta m^2_{21} - \Delta \bar m^2_{21}|$ & 1.1$\times 10^{-4}\,
     \mathrm{eV^2}$& 2.2$\times10^{-5}\,\mathrm{eV^2}$ \\
\hline \rule{0pt}{2pt}&\\
       \strut$|\Delta m^2_{31} - \Delta \bar m^2_{31}|$ & 1$\times 10^{-2}\,
     \mathrm{eV^2}$& 3.3$\times10^{-5}\,\mathrm{eV^2}$  \\
\hline
\end{tabular}
}\end{center}
\vspace*{-0.2cm}

In connection with difficulties arisen from interpreting the entire
body of experimental results on NO, even resorting to `marginal'
solutions with the sterile neutrinos (see, e. g., Refs
 \cite{str}, \cite{MALT}) \footnote{The SNO experiments data
eliminate the need to consider sterile neutrinos (\cite{SNO} and \cite{Ba}); for the subsequent estimates
 see  paper \cite{ster}, talk \cite{STR} and  review \cite{MSTV}; see the situation after the MiniBooNE first result, e.g.,
 in very recent works \cite{ChDG}.}, an extended set of squared mass differences was used
(via an independent value of $\Delta \bar m^2_{ij}$ for the
antineutrino); to justify this usage, its origin is connected  to the
hypothetical CPT violation in the neutrino sector of the theory via $m_{\bar \nu}\neq ~m_{\nu}$. It
should be emphasize that all models involved \cite{mur}, \cite{BL,bar2,bar1} lack theoretical foundation: in fact, the
difference between the neutrino and antineutrino masses is
introduced into them `by hand' since CPT violation is impossible in
a Lorentz-invariant theory  (see \cite{gre, gr2}) \footnote  {An
obvious assumption is made of   the usual connection between spin
and statistics.}  as we stressed in Section 2.

Murayama and Yanagida \cite{mur} were the first to try this approach
to interpret  experimental results on NO and supernova neutrino events (see also \cite{Ahl}).
By analyzing the energies of neutrino events in the Kamiokande and IMB experiments
which  originated from the SN1987A,  these authors obtained
arguments against preferred values of $\Delta m^2_{\rm LSND}\thickapprox0.1 - 1~$eV$^2$.
Then they proposed a scheme of neutrino and antineutrino masses
that is compatible with all neutrino oscillations data and the LSND anomaly without adding the sterile neutrino.
Furthermore, characteristics of above neutrino events  do not contradict this scheme too. The LSND result is
interpreted as a consequence of assuming large squared mass difference for the antineutrino, $\Delta \bar m^2$.
The authors considered it essential to test for this assumption in the MiniBooNE experiment with an antineutrino beam.

The situation with interpretation of NO data on the basis of CPT non-invariant neutrino mass spectra
is outlined in Ref. \cite{GMS} and in   review talks  \cite{Sch,Schw}. An analysis of all results, with
the LSND experiment either taken or not taken into account, was carried out in  the three-flavor approach.\\
It was shown that areas allowed for the LSND and for all other experiments on the $\Delta m^2$~$vs$~ sin$^{2}2\theta$
 plane do not overlap at the 3$\sigma$
level  while the values of these parameters that correspond to the best fit are practically identical
 in scenarios with and without CPT violation. The obtained level for rejecting the interpretation with CPT violation
 is 4.6$\sigma$  \cite{Schw}.

After the recent global  fitting \cite{GMS,GGMM} of all the data for NO  except for the LSND result, the current
 values of mixing parameters were obtained separately for neutrinos and antineutrinos. The resultant picture of
a possible scenario with CPT violation is illustrated at this page by Fig.\,55 (four joined panels are due to
absence of information about the relative ordering of neutrinos $vs$ antineutrinos states),
which is taken from  Ref.\,\cite{GGMM}. It turnes to be impossible to tell  CPT-violating scenarios with normally
 ordered $\nu$ states plus inversely ordered $\bar\nu$ states (or vice versa)  from the CPT-conserving one.
 The analysis shows also \cite{GMS,GGMM} that the CPT-violating scenario of Ref. \cite{bar1} fails to give
 a good description simultaneously of both the LSND result and the all-but-LSND NO data.\\

\begin{figure}
\begin{center}
\vspace*{-1.6cm}
\includegraphics[bb=0 0 567 841,width=16cm]{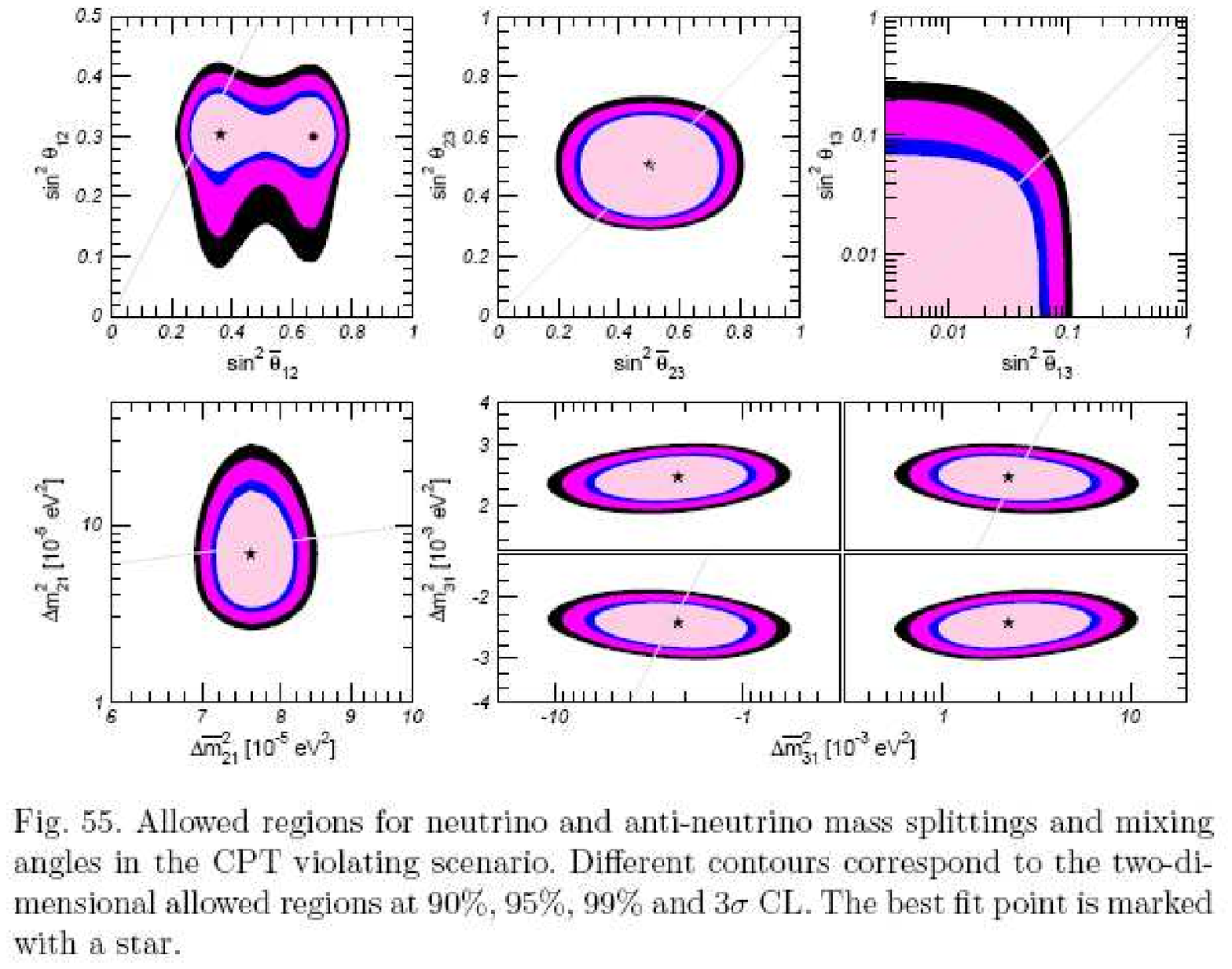}
\vspace*{-13cm}
\end{center}
\end{figure}



\noindent {\large\bf  Conclusion}\\[-10pt]

Contents of this article characterize the current
state of the old problem of theoretical and experimental
investigation of the hypothetical violation of the CPT and/or Lorentz invariance
within the neutrino sector of the Standard Model. The information presented gives us
evidence for constituting the neutrino oscillations as a novel area in
which promising possibilities are opened for testing this symmetries. One should not forget, however, the
status of the CPT invariance as one of the fundamental concepts underpinning the theory.

 This is a proper place to quote the following outlook which is undoubtedly shared by most of the
researchers: \textit{Of course, whatever could be measured should be
measured and whatever could be tested should be tested. There should
be no reservations: such a fundamental symmetry as CPT should be
tested. However one should keep in mind that unlike breaking of C,
P, T, CP, PT, and TC, the breaking of CPT is non-compatible with the
standard quantum field theory, the only basis for a self-consistent
phenomenological description of any process, which we know up to
now. Therefore the chances that CPT breaking would be discovered are
vanishingly small.}$~$\cite{ok2}

Indeed, searches for CPT violation are unsuccessful yet (see Section 2), even in neutrino physics,
as one discusses in Section 3 -- 5.

No evidence for Lorentz violation  was found too,
in spite of plenty of theoretical studies and
 advances in  the precision testing which are carried out over the last decade or two.
Hence, it is natural to start a prudent outlook of the wise expert with the question:
\textit{...
when have we tested enough? We currently have bounds on Lorentz violation strong enough that
there is no way to put Lorentz violating operators of dimension $\leq$ 6 coming solely from
Planck scale physics into our field theories. It therefore seems hard to believe that Lorentz
invariance  could be  to restrict the classes of quantum gravity theories/spacetime models
we should consider. Without a positive signal of Lorentz violation, this is all that can
reasonably be hoped for.} \cite{matt}

On the other hand, from the standpoint of quantum gravity (QG),
as the future theory to which theoreticians address as a physical origin of both CPT and
Lorentz violation (LV), true adherents of QG believe that \textit{...
LV is not the only possible low energy QG signature. Nonetheless, it
is encouraging that it was possible to gather such strong
constraints on this phenomenology in only a few years. This should
motivate researches to further explore this possibility as well as
to look even harder for new QG induced phenomena that will be
amenable to observational tests. This will not be an easy task, but the
data so far obtained prove that the Planck scale is not untestable
after all.} \cite{Lib}
\\[-25pt]

\end{document}